\documentclass[12pt, onecolumn]{IEEEtran} 
\usepackage{microtype}
\usepackage{subfigure}
\usepackage{epsfig, color}
\usepackage{amsmath}
\usepackage{amsthm}
\usepackage{amssymb}
\usepackage{multirow}
\usepackage{booktabs}

\usepackage{mathtools}
\usepackage{float}
\usepackage{comment}
\usepackage[ruled,vlined]{algorithm2e}
\usepackage{setspace}
\SetArgSty{textup}
\usepackage[para]{footmisc}
\usepackage{cite}

\restylefloat{table}
\usepackage[
  separate-uncertainty = true,
  multi-part-units = repeat
]{siunitx}

\begin{document}

\title{High Dimensional Channel Estimation Using Deep Generative Networks}
\author{{
    }	Eren Balevi, Akash Doshi, Ajil Jalal, Alexandros Dimakis, Jeffrey G. Andrews
\thanks{The authors are with the University of Texas at Austin, TX, USA. Contact Author Email: jandrews@ece.utexas.edu. This work was supported in part by Intel. This paper was presented in part at the $21^{st}$ IEEE Signal Processing Advances in Wireless Communications Workshop, May 2020 in the Special Session for Machine Learning in Communications \cite{Dos20BalAnd}.	}	
}

\maketitle 
\normalsize
\begin{abstract}
This paper presents a novel compressed sensing (CS) approach to high dimensional wireless channel estimation by optimizing the input to a deep generative network. Channel estimation using generative networks relies on the assumption that the reconstructed channel lies in the range of a generative model. Channel reconstruction using generative priors outperforms conventional CS techniques and requires fewer pilots. It also eliminates the need of a priori knowledge of the sparsifying basis, instead using the structure captured by the deep generative model as a prior. Using this prior, we also perform channel estimation from one-bit quantized pilot measurements, and propose a novel optimization objective function that attempts to maximize the correlation between the received signal and the generator's channel estimate while minimizing the rank of the channel estimate. Our approach significantly outperforms sparse signal recovery methods such as Orthogonal Matching Pursuit (OMP) and Approximate Message Passing (AMP) algorithms such as EM-GM-AMP for narrowband mmWave channel reconstruction, and its execution time is not noticeably affected by the increase in the number of received pilot symbols.
\end{abstract}


\section{Introduction}

\subsection{Motivation}
To meet the demand for extremely high bit rates and much lower energy consumption per bit, future wireless systems are trending to bandwidths larger than 1 GHz and carrier frequencies above 100 GHz. As an example, a future communication system (6G and beyond) may operate at a carrier frequency approaching 300 GHz with well over 10,000 cross-polarized antenna elements at each transceiver, and antenna spacings on the order of 1-2 mm \cite{rappaport2019wireless, elayan2018terahertz}. For channel estimation in many antenna systems, typically the number of pilots is assumed to be larger than the number of transmit antennas, resulting in significant training overhead which does not scale well to such high dimensional future communication systems. Using sparsity with a compressed sensing method to alleviate this problem leads to solving a complex optimization problem at every coherence interval, whose complexity scales with the number of antennas, will become infeasible. Thus, existing approaches to channel estimation will not scale to this regime in terms of complexity, power consumption, or pilot overhead, and fundamentally new methods are needed. The key to simplifying channel estimation in such high dimensional systems is to exploit stronger prior knowledge of the channel  structure. In this paper we propose a novel unsupervised learning-based approach using deep generative networks for channel estimation. 

\subsection{Related Work}
Traditional training-based channel estimators such as least-squares (LS) are optimum Maximum Likelihood estimators for rich multipath channels. Furthermore, for Gaussian signal recovery with a known correlation matrix, minimum mean-squared estimators (MMSE) find the signal estimate $x$ that maximizes the a posteriori probability $p(x|y)$ and outperforms LS \cite{bjornson2017massive}. However recent channel measurements conducted for mmWave and THz cellular systems have indicated that, due to clustering of the paths into small, relatively narrowbeam clusters, high dimensional channels are often very sparse in their beamspace representation \cite{rappaport2019wireless} or their spatial covariance matrix is low rank \cite{eliasi2017low}. Among the first papers to highlight the need for exploiting these sparse structures, that LS and MMSE cannot exploit, was \cite{bajwa2010compressed}, which exploited channel sparsity in the beamspace representation of a multi-antenna channel to formulate channel estimation as a CS problem, while \cite{bajwa2009sparse} also highlighted how to exploit sparsity in the delay-Doppler domain. MmWave channel estimation is made difficult by the low received SNR due to high omnidirectional path loss, and to combat this path loss, large antenna arrays are used to obtain beamforming gain. In \cite{alkhateeb2014channel}, a sparse formulation of the mmWave channel estimation problem was given by expressing the sensing matrix $\Psi$ as a function of the transmit and receive antenna array response vectors, in addition to the training precoders and combiners. An open loop strategy for downlink mmWave channel estimation and design of precoders/combiners that minimize the coherence of $\Psi$, while incorporating hybrid constraints at the transceiver, was presented in \cite{mendez2016hybrid}, enabling reconstruction from a small number of measurements. In \cite{alkhateeb2014channel} and \cite{mendez2016hybrid}, Orthogonal Matching Pursuit ($\mathcal{L}0$ norm minimization) and Basis Pursuit Denoising ($\mathcal{L}1$ norm minimization) were employed for sparse channel reconstruction. Approximate Message Passing (AMP) is another robust class of techniques for compressed sensing \cite{rangan2011generalized}, and variants such as EM-GM-AMP \cite{vila2013expectation} and VAMP \cite{rangan2019vector} outperform OMP and BPDN for a large class of sensing matrices. AMP has been widely advocated for MIMO channel estimation in the research community, especially for low resolution receivers \cite{mo2014channel,mo2017channel}. AMP has even been extended to adaptively learn the clustered structure in the angle-delay domain in \cite{lin2018estimation}.

However, real world channels are never exactly sparse in the DFT-basis, nor do we know the basis that would yield the most sparse representation. Moreover, all these techniques involve solving a complex optimization problem at each interval, and require a large number of pilots, especially in low resolution receivers. These are some of the reasons why CS-based methods are still not employed in conventional WiFi receivers for channel estimation, which typically employ LS channel estimation with frequency domain smoothing that leverages the coherence bandwidth of the channel \cite{Freq_Smoothing}.

Meanwhile, there has been a rapid advancement in the application of techniques from deep learning to channel estimation for massive MIMO and mmWave systems. One of the approaches taken was to perform Joint Channel Estimation and Signal Detection (JCESD) \cite{YeLi18} \cite{HeLi19}, thus performing channel estimation implicitly. Not recovering the channel estimate prevents precoder and/or combiner optimization, and these techniques call for extensive signal processing changes at the transceiver. One obvious way to recover the channel estimate is to train a Neural Network (NN) in a supervised manner, such that it is trained to take as input the pilot measurements and output the channel matrix. This approach is taken in many recent papers \cite{yang2019deep,He18Li,ru2019model,Dong19Gaspar,GaoLi18}. In particular, \cite{yang2019deep} also appends the LS channel estimate of the current and previous received pilot signal to the NN's input to improve its performance. In \cite{He18Li}, a variant of the AMP technique called LDAMP is unfolded into a NN, by making the parameters of LDAMP learnable. Exploiting the inherent structure in a spatial channel matrix, making its estimation analogous to image reconstruction, \cite{ru2019model} and \cite{Dong19Gaspar} employ Convolutional Neural Networks (CNNs) in place of Fully Connected NNs to learn a channel estimator. A novel refinement called SIP-DNN was proposed in \cite{GaoLi18}, that chose to estimate the channel at all the receive antennas using only the signal received by the high-resolution ADC antennas. 

However building such labeled channel datasets for a supervised task is time-consuming, and most of these techniques would not perform well if the received signal was corrupted by hardware impairments and/or transient effects such as shadow fading. A few papers recently have been using techniques from unsupervised learning to overcome this limitation of having to build a huge labeled dataset. In \cite{BalAndDCE} and \cite{8949757}, they combine an LS estimator with an underparameterized CNN-based denoiser called Deep Decoder \cite{Heckel18} to exploit correlation in the channel estimate to improve its quality. In \cite{wen2018deep}, they train an autoencoder to learn a compressed representation for the channel that could immensely reduce channel state information (CSI) feedback overhead in massive MIMO, while \cite{lu2019multi} trains a CRNet with the same objective in multi-resolution receivers.

\subsection{Contributions}
In summary, most of the proposed Deep Learning techniques are discriminative, meaning that a priori information is not exploited as opposed to generative models and often call for drastic changes in transceiver signal processing, while the existing signal processing techniques are designed for sparse signal recovery. As mentioned before, there is no explicit way to determine the basis that will generate the sparse channel representation with the least non-zero entries, which would allow perfect recovery for a wider range of sensing matrices with the same number of measurements. This is where compressed sensing using generative models proves useful. By finding an approximate solution in the span of a generative model, \cite{bora2017compressed} shows how to achieve CS guarantees without employing sparsity. The authors of \cite{bora2017compressed} present a simple gradient descent based algorithm that enables signal recovery for inherently sparse or structured signals from compressed measurements by exploiting the prior learnt by a generative model. In this paper, we draw inspiration from the approach presented in \cite{bora2017compressed} to perform the estimation of high dimensional wireless  channels from compressive pilot measurements. Our contributions are elaborated below:  

\textbf{Training a GAN to learn the channel distribution:} The underlying probability distribution of spatial channel matrices for a particular environment can be very complex, and analytically intractable. We describe how to train a Wasserstein GAN \cite{arjovsky2017wasserstein} using a set of simulated channel realizations, such that it learns a generator model that is capable of drawing samples from the underlying channel distribution. 

\textbf{Full resolution channel estimation:} The trained generative model will output channel realizations for different input vectors. We describe a procedure to find the optimal input vector such that we can use the prior of the trained generator to find the channel estimate from a low number of noisy pilot measurements. Moreover, the optimization problem defined by the generative network operates in a low-dimensional subspace, whose dimensionality is independent of the number of received pilots, and achieves significant reduction in computational complexity. Simultaneously, our technique also helps to develop a channel representation that drastically reduces CSI feedback overhead, and learn a prior that enables it to significantly outperform conventional CS techniques, without knowledge of the sparsifying basis. 

\textbf{One-bit quantized channel estimation:} We design a custom loss function that aims to find the channel estimate, in the range of the generator's output, that has low rank while maximizing correlation with the received one-bit measurements. We compare its performance with state-of-the-art CS techniques such as EM-GM-AMP \cite{vila2013expectation}, and find that it significantly improves the quality of the channel estimate, while still requiring only a limited number of pilots. We validate the improvement in the channel estimate by evaluating the throughput for a hybrid precoded data transmission, where the RF and baseband precoders were designed using OMP \cite{el2014spatially}.

The paper is organized as follows. The system model is outlined in Section \ref{sec:system_model}. The generative channel estimator is explained in detail in Section \ref{sec:gce}, the NN architecture details, simulation benchmarks and results are outlined in Section \ref{sec:Simulations} and \ref{sec:results} respectively, and the conclusions are highlighted in Section \ref{sec:conc}.

\section{System Model} \label{sec:system_model}

\subsection{Training based channel estimation} \label{subsec:tce}
Consider a point-to-point downlink (DL) MIMO setup, where the base station is equipped with $N_{\mathrm{t}}$ transmit antennas and the User Equipment (UE) is equipped with $N_{\mathrm{r}}$ receive antennas. For simplicity, the exposition that follows considers only a single narrowband frequency channel but can easily be extended to multiple ($N_\mathrm{f} > 1$) subcarriers. We consider hybrid beamformers and combiners, and present a training-based channel estimation approach.

In the DL channel estimation phase, the BS uses a training beamformer $\mathbf{p} \in \mathbb{C}^{N_{\mathrm{t}} \times 1}$ to transmit a symbol $s \in \mathbb{C}$. To simplify analysis, we set $s = 1$ in all experiments, but retain it in the equations for ease of understanding. The UE employs $N_{\mathrm{r}}$ RF chains, hence for each beamforming vector $\mathbf{p}$, $N_{\mathrm{r}}$ measurements are produced at the UE. We assume that the training combiner $\mathbf{q}_i \in \mathbb{C}^{N_{\mathrm{r}} \times 1}$ $i \in [N_{\mathrm{r}}]$ is a 1-sparse vector with 1 at the $i^{\mathrm{th}}$ position. 
As explained in \cite{mendez2016hybrid}, the number of measurements per time instant at the UE does not depend on the number of RF chains employed at the BS. The total number of measurements $M = N_{\mathrm{r}}N_{\mathrm{p}}$ where $N_{\mathrm{p}}$ is number of distinct beamforming vectors $\mathbf{p}$ employed by the BS during training. We denote this sequence as $\mathbf{P} = [\mathbf{p}_1 ... \mathbf{p}_{N_{\mathrm{p}}}] \in \mathbb{C}^{N_{\mathrm{t}} \times N_{\mathrm{p}}}$. It is assumed that the channel coherence time is greater than $N_{\mathrm{p}}T$, where $T$ is the symbol period, hence the spatial channel matrix $\mathbf{H} \in \mathbb{C}^{N_{\mathrm{r}} \times N_{\mathrm{t}}}$ remains constant over the $N_{\mathrm{p}}$ time slots. Hence the received training signal $\mathbf{Y} \in \mathbb{C}^{N_{\mathrm{r}} \times N_{\mathrm{p}}}$ at the UE can be written as \cite{alkhateeb2014channel}
\begin{equation} \label{eq:rec_signal}
    \mathbf{Y} = \mathbf{HP}s + \mathbf{N},
\end{equation}
where each element of $\mathbf{N} \in \mathbb{C}^{N_{\mathrm{r}} \times N_{\mathrm{p}}}$ are independent and identically distributed complex Gaussian random variables with mean 0 and variance $\sigma^2$. To have more compact expressions, the matrices are defined as vectors by concatenating the columns, yielding
\begin{equation} \label{eq:rec_signal_vec}
    \underline{\mathbf{y}} = \underline{\mathbf{HP}}s + \underline{\mathbf{n}},
\end{equation}
where $\underline{\mathbf{y}}, \underline{\mathbf{HP}}, \underline{\mathbf{n}} \in \mathbb{C}^{N_{\mathrm{r}}N_{\mathrm{p}} \times 1}$. Writing $\mathbf{HP}$ as $\mathbf{I_{N_{\mathrm{r}}}HP}$, and utilizing the expansion $\underline{\mathbf{ABC}} = (\mathbf{C}^{T} \otimes \mathbf{A})\underline{\mathbf{B}}$, we have 
\begin{equation} \label{eq:rec_signal_kron}
    \underline{\mathbf{y}} = (\mathbf{P}^{T} \otimes \mathbf{I_{N_{\mathrm{r}}}})\underline{\mathbf{H}}s + \underline{\mathbf{n}},
\end{equation}
 where $T$ denotes the transpose operator and $\otimes$ denotes the Kronecker product. Clearly, the system of equations represented by \eqref{eq:rec_signal_kron} does not have a unique solution if $N_{\mathrm{p}} < N_{\mathrm{t}}$. In other words, the LS channel estimate $\mathbf{\hat{H}}$ given by
\begin{equation} \label{eq:ls_eq_rep}
    \mathbf{\hat{\underline{H}}} = \underset{\mathbf{\underline{H}} \in \mathbb{C}^{N_{\mathrm{r}}N_{\mathrm{t}} \times 1}}{\rm argmin} \hspace{0.05 in} ||\underline{\mathbf{y}} - (\mathbf{P}^{T} \otimes \mathbf{I_{N_{\mathrm{r}}}})\underline{\mathbf{H}}s||^2
\end{equation}
has multiple solutions. Thus, in the low density pilot regime, one cannot directly use LS channel estimation. If $\mathbf{H}$ is inherently sparse or structured in a known basis, this can be exploited by CS algorithms, and is explained as part of the baselines for comparison in Section \ref{subsec:baselines}. The above notation also extends easily to the case where the received signal is quantized, with \eqref{eq:rec_signal_kron} being rewritten as 
\begin{equation} \label{eq:rec_signal_kron_q}
    \underline{\mathbf{y}} = \mathcal{Q}_\mathrm{n}((\mathbf{P}^{T} \otimes \mathbf{I_{N_{\mathrm{r}}}})\underline{\mathbf{H}}s + \underline{\mathbf{n}}),
\end{equation}
where $\mathcal{Q}_\mathrm{n}$ denotes the $\mathrm{n}$ bit quantization operator.

\subsection{Hybrid Precoding for Data Transmission} 
In Section \ref{subsec:tce}, we presented a training-based channel estimation approach, hence the training beamformers used were random sequences of QPSK symbols (one can also use a random subset of the columns of the DFT matrix or the Zadoff-Chu sequences). Now we move from the training stage to the data transmission phase, where to obtain a higher throughput, one performs optimization of the precoder matrices $\mathbf{F}_{\mathrm{RF}}$ and $\mathbf{F}_{\mathrm{BB}}$ at the BS, in which $\mathbf{P}=\mathbf{F}_{\mathrm{RF}}\mathbf{F}_{\mathrm{BB}}$. To achieve this, the channel estimate recovered at the UE is conveyed to the BS, to maximize the information-theoretic capacity, while incorporating the hardware and power constraints imposed on the entries of $\mathbf{F}_{\mathrm{RF}}$ and $\mathbf{F}_{\mathrm{BB}}$. As outlined in \cite{el2014spatially}, 
we utilize spatially sparse precoding via Orthogonal Matching Pursuit to find the optimum $\mathbf{F}_{\mathrm{RF}}$ and $\mathbf{F}_{\mathrm{BB}}$ and evaluate the throughput.

\section{Generative Channel Estimator} \label{sec:gce}
\subsection{Training a GAN to learn the channel distribution}
We use Generative Adversarial Networks (GANs) for training a generative model. Despite the extensive recent application of deep learning to wireless communications, few communication papers have employed GANs, owing to their perceived training instability \cite{srivastava2017veegan}. In \cite{o2019approximating}, the authors proposed the use of variational GANs to accurately learn the channel distribution. However, they restricted themselves to additive noise, and did not consider fading or MIMO. In \cite{ye2018channel}, they employ a conditional GAN that is trained to output the received signal when the transmitted signal and the received pilot information is appended to the input of the GAN. However, when extending it to fading channels, they assumed that the real channel response was available as input to the GAN. Moreover, none of these papers exploit the compressed representation that the generator of a GAN learns for a given output signal. We now give an overview of the training procedure for GAN's in the context of spatial channel matrix generation.

A GAN \cite{goodfellow2014generative} consists of two feed-forward neural networks, a generator $\mathbf{G}(z;\theta_g)$ and a discriminator $\mathbf{D}(x;\theta_d)$ engaging in an iterative two-player minimax game with the value function $V(\mathbf{G},\mathbf{D})$:
\begin{equation}\label{eq:GAN_loss_fn}
    \underset{\mathbf{G}}{\operatorname{min}}~\underset{\mathbf{D}}{\operatorname{max}} ~V(\mathbf{G},\mathbf{D}) =  \mathbb{E}_{x\sim \mathbb{P}_{r}(x)}[h_{\mathbf{D}}(\mathbf{D}(x;\theta_d))] 
    + \mathbb{E}_{z\sim \mathbb{P}_{z}(z)}[h_{\mathbf{G}}(\mathbf{D}(\mathbf{G}(z;\theta_g);\theta_d))],
\end{equation}
where $\mathbf{G}(z)$ represents a mapping from the input noise variable $z\sim \mathbb{P}_{z}(z)$ to the data space $x\sim \mathbb{P}_{r}(x)$, while $\mathbf{D}(x)$ represents the probability that $x$ came from the data rather than $\mathbf{G}$. The exact form of $h(.)$ depends on the choice of loss function. In \cite{goodfellow2014generative}, $h_{\mathbf{D}}(\mathbf{D}(x)) = \log{\mathbf{D}(x)}$ whereas $h_{\mathbf{G}}(\mathbf{D}(\mathbf{G}(z))) = \log{(1-\mathbf{D}(\mathbf{G}(z)))}$. On the other hand, in the Wasserstein GAN proposed in \cite{arjovsky2017wasserstein}, $h_{\mathbf{D}}(\mathbf{D}(x)) = \mathbf{D}(x)$ and $h_{\mathbf{G}}(\mathbf{D}(\mathbf{G}(z))) = -\mathbf{D}(\mathbf{G}(z))$. Given $z \in \mathbb{R}^d$ and $\mathbf{G}(z) \in \mathbb{R}^n$, typically $z \sim \mathcal{N}(\mathbf{0},\mathbf{I}_d)$ and $d \ll n$. For example, when a GAN is trained on an image dataset, $d$ can be 100, while $n = 64\times64\times3 = 12288$ (where 64 represents the image height and width in pixels and 3 represents the RGB color triplet). $\mathbf{G}$ is said to implicitly learn the distribution $\mathbb{P}_{g}$ (stored in its weights $\theta_g$), which on convergence, should approach $\mathbb{P}_{r}$.

\begin{figure}
    \centering
    \includegraphics[width=5in]{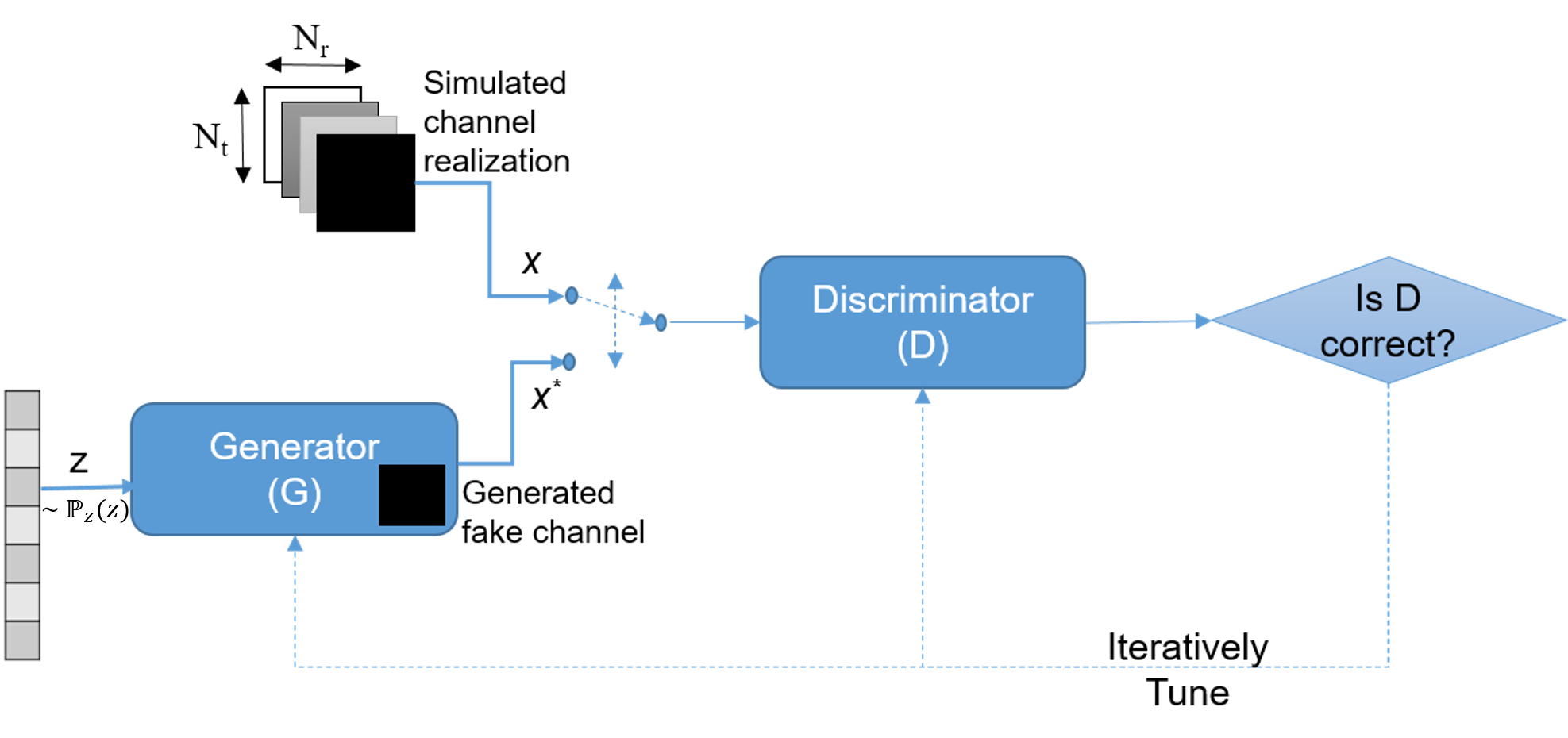}
    \caption{Training a GAN for spatial channel matrices}
    \label{fig:gan_training}
\end{figure}

Since the seminal paper \cite{goodfellow2014generative}, numerous variants of GAN have been published, differing in the  architecture and/or training procedure of $\mathbf{G}$ and $\mathbf{D}$ or the loss function used for penalizing the output of $\mathbf{D}$ \cite{radford2015unsupervised}, \cite{arjovsky2017wasserstein}. However, GANs are known to be difficult to train, one of the reasons being that they are subject to mode collapse. That is, they learn to characterize only a few modes of the distribution \cite{srivastava2017veegan}. The objective of training a GAN is that by varying the weights $\theta_g$ and $\theta_d$ of $\mathbf{G}(z;\theta_g)$ and $\mathbf{D}(z;\theta_d)$, we want $\mathbb{P}_g \rightarrow \mathbb{P}_r$. In \cite{arjovsky2017wasserstein}, the Wasserstein-1 (EM) distance is shown to be much weaker than KL (Kullback-Leibler) or JS (Jensen-Shannon) divergences\footnote{A set of probability distributions $\mathbb{P}_n$ is said to converge to $\mathbb{P}_\infty$ under a distance metric $\rho$ if $\rho(\mathbb{P}_n,\mathbb{P}_\infty) \rightarrow 0$ as $n \rightarrow \infty$. By ``weaker", we mean that the set of convergent sequences under EM is a superset of the sequences convergent under KL or JS}, such that simple sequences of probability distributions converge under EM but not KL or JS. Using the continuous and differentiable EM distance as the loss function for the output of $\mathbf{D}$ during training with weights clipping eliminates careful balance in training of $\mathbf{D}$ and $\mathbf{G}$, and design of NN architecture. It also drastically reduces mode collapse since we can train $\mathbf{D}$ to optimality. Hence in this paper, we employ the Wasserstein GAN \cite{arjovsky2017wasserstein} for learning the spatial channel distribution. An outline of the procedure for training a Wasserstein GAN in the context of spatial channel matrix generation is given in Alg. \ref{alg:GAN_training} (adapted from \cite{arjovsky2017wasserstein}) \footnotetext{The original paper \cite{arjovsky2017wasserstein} refers to the discriminator as critic and uses $n_{\mathrm{critic}} = 5$ which we refer to as $n_{\mathrm{d}}$} and depicted in Fig. \ref{fig:gan_training}.

\begin{algorithm} \label{alg:GAN_training}
\setstretch{1}
\SetAlgoNoLine
 $\mathbf{D}$ should output 1 for a true channel realization $x\sim \mathbb{P}_{r}(x)$ and 0 for a generated fake \\channel realization $\mathbf{G}(z) \sim \mathbb{P}_{g}$ when $z$ is sampled from $\mathbb{P}_{z}$ \\
 \For{number of training iterations}{
 \For{$n_{\mathrm{d}}$ iterations}{
  \begin{itemize}
      \item Sample minibatch of $m$ noise samples $\{z_1, ..., z_m\} \sim \mathbb{P}_z$. Update $\mathbf{D}$ by \\ascending its stochastic gradient 
      \begin{equation*}
          \nabla_{\theta_d} \frac{1}{m} \sum_{i=1}^{m} -\mathbf{D}(\mathbf{G}(z_i)))
      \end{equation*}
      \item Sample minibatch of $m$ channel realizations $\{x_1, ..., x_m\} \sim \mathbb{P}_r$. Update $\mathbf{D}$ by \\ascending its stochastic gradient 
      \begin{equation*}
          \nabla_{\theta_d} \frac{1}{m} \sum_{i=1}^{m} \mathbf{D}(x_i)
      \end{equation*}
      \item $\theta_d$ = clip$(\theta_d,-c,c)$
      \end{itemize}}
      Sample minibatch of $m$ noise samples $\{z_1, ..., z_m\} \sim \mathbb{P}_z$. Update $\mathbf{G}$ by \\descending its stochastic gradient
      \begin{equation*}
          \nabla_{\theta_g} \frac{1}{m} \sum_{i=1}^{m} -\mathbf{D}(\mathbf{G}(z_i)))
      \end{equation*}
 }
\caption[caption]{Minibatch stochastic gradient descent training of Wasserstein GANs for spatial channel matrix generation with $n_{\mathrm{d}} = 5$ and $c = 0.01$}
\end{algorithm}

\subsection{CS-based channel estimation using generative networks}
Consider a noisy compressive measurement $\mathbf{y}$ of an image $\mathbf{x}^*$ such that $\mathbf{y} = \mathbf{Ax}^* + \mathbf{n}$. A simple gradient descent based approach for compressed sensing using generative networks was proposed in \cite{bora2017compressed} to find the low dimensional representation $z^*$ of the given input image $\mathbf{x}^*$ such that the reconstructed image $\mathbf{G}(z^*)$ has small measurement error $||\mathbf{y} - \mathbf{AG}(z)||_2^2$. While this is a non-convex objective to optimize (since $\mathbf{G}(z)$ is a non-convex function of $z$), gradient descent was found empirically to work well. To reconstruct the image, \cite{bora2017compressed} solves the following optimization problem:
\begin{equation}\label{eq:main_opt_prb}
    z^* = \underset{z} {\mathrm{arg\ min\ }} f(\mathbf{y},\mathbf{AG}(z)),
\end{equation}
where $\mathbf{y}$ is the vector of received samples, $\mathbf{G}$ is a generative model, $\mathbf{A}$ is a measurement matrix, and $f$ is a loss function. For example, we could have $f(\mathbf{y},\mathbf{AG}(z)) = ||\mathbf{y} - \mathbf{AG}(z)||_2^2$. Here, we minimize the loss function over the input variable to the generator $z$. The reconstructed image is then $\mathbf{G}(z^*)$. As long as gradient descent finds a good approximate solution to \eqref{eq:main_opt_prb}, \cite{bora2017compressed} gives a theoretical proof to show that $\mathbf{G}(z^*)$ will be almost as close to the true $\mathbf{x}^*$ as the closest possible point in the range of $\mathbf{G}$, when the entries of the sensing matrix $\mathbf{A}$ are sub-Gaussian\footnote{A random variable $X \in \mathbb{R}$ is said to be sub-Gaussian with variance-proxy $\sigma^2$ if $\mathbb{E}[X] = 0$ and its moment generating function satisfies $\mathbb{E}[\exp(sX)] \leq \exp(\sigma^2s^2/2)$ for all $s \in \mathbb{R}$}. 

\begin{figure}
    \centering
    \includegraphics[width=5in]{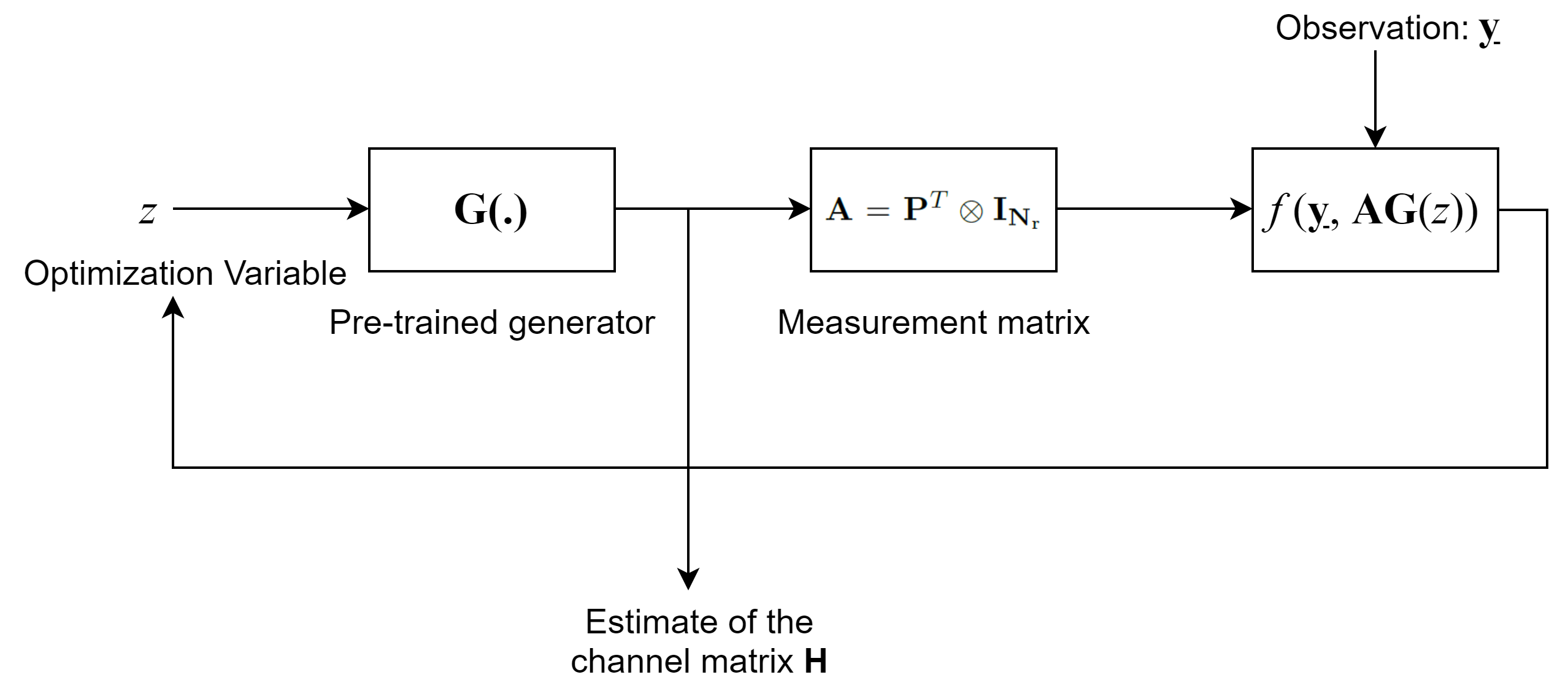}
    \caption{Generative Channel Estimator Framework}
    \label{fig:gan_framework}
\end{figure}

To adapt the framework presented in \cite{bora2017compressed} for channel estimation, we first train a Wasserstein GAN \cite{arjovsky2017wasserstein} using a set of realistic channel realizations $\mathbf{H}$ (details of channel parameters presented in Section \ref{sec:Simulations}) as defined in \eqref{eq:rec_signal}. We then extract the trained generator $\mathbf{G}$. The trained generator, having implicitly learned the underlying probability distribution of the channel matrices, will output channel realizations $\mathbf{G}(z)$ for a given $\mathcal{L}2$ bounded input vector $z$. In the testing phase, we will be given the noisy pilot measurements $\mathbf{\underline{y}}$ as defined in \eqref{eq:rec_signal_kron}. We consider two possible cases: when the measurements are full resolution and when they are one-bit quantized. For each case, we have heuristically developed loss functions, that define the optimization problem to be solved at every coherence interval using gradient descent. An illustration of the framework is shown in Fig \ref{fig:gan_framework} and the approach is summarized in Alg. \ref{alg:CS_GAN}. We refer to this framework as the Generative Channel Estimator (GCE).
\textbf{Full Resolution Channel Estimation: } Replacing the sensing matrix $\mathbf{A}$ by $\mathbf{P}^{T} \otimes \mathbf{I_{N_{\mathrm{r}}}}$ as derived in \eqref{eq:rec_signal_kron}, and imposing an $\mathcal{L}2$ bound on $z$ via regularization, we attempt to solve the following non-convex optimization problem:
\begin{equation} \label{eq:gan_eq_rep}
    z^* = \underset{z \in \mathbb{R}^d} {\mathrm{arg\ min\ }} \hspace{0.05 in} ||\underline{\mathbf{y}} - (\mathbf{P}^{T} \otimes \mathbf{I_{N_{\mathrm{r}}}})\mathbf{\underline{G}}(z)s||_{2}^2 + \lambda_{\mathrm{reg}} ||z||_{2}^2,
\end{equation}
where $d$ is the dimension of the input vector to the GAN and $\lambda_{\mathrm{reg}}$ serves as a regularization parameter. The reconstructed channel estimate is then simply $\mathbf{G}(z^*)$. Note that the entries in the training precoder $\mathbf{P}$ were chosen i.i.d. from QPSK symbols. As a consequence, all the entries of the matrix $\mathbf{A} = \mathbf{P}^{T} \otimes \mathbf{I_{N_{\mathrm{r}}}}$ are bounded (being either 0 or QPSK symbols) with mean 0, and from Hoeffding's Lemma \footnote{Hoeffding's Lemma states that for any random variable $X$ with $\mathbb{E}[X] = 0$ such that $a \leq X \leq b$ w.p. 1, for all $s \in \mathbb{R}$, $\mathbb{E}[\exp(sX)] \leq \exp(s^2(b-a)^2/8)$. Hence $X$ is sub-Gaussian with variance proxy $(b-a)^2/4$.} applied separately to the real and imaginary parts, it follows that each entry of A will be sub-Gaussian.

\textbf{Quantized Channel Estimation: } We now consider the case where the received signal is 1-bit quantized. As a result, MIMO channel estimation even in the noiseless setting with sufficient pilot symbols is an under-determined problem. In \cite{myers2019low}, they exploit the low-rank nature of mmWave channels (due to clustering in the propagation environment) to constrain the space of channel estimates to matrices $\mathbf{H}$ with low nuclear norm $||\mathbf{H}||_*$ (a relaxation of the low-rank constraint). In \cite{Qiu19WeiQiu}, the authors solve the same optimization problem as \eqref{eq:main_opt_prb} with the measurements $\mathbf{y}$ being one-bit quantized, and under certain assumptions on the measurement matrix ($\mathbf{A}$ in \eqref{eq:main_opt_prb}) and the architecture of the GAN, design a custom loss function to solve for $z^*$. We draw inspiration from the approach taken in \cite{myers2019low} and \cite{Qiu19WeiQiu} to design the following non-convex optimization problem for recovery in one-bit setting:
\begin{equation} \label{eq:gan_eq_rep_obq}
    z^* = \underset{z \in \mathbb{R}^d} {\mathrm{arg\ min\ }} \hspace{0.05 in} -\lambda_{\mathrm{reg}} \sum_{\mathrm{i}=1}^{N_{\mathrm{p}}N_{\mathrm{r}}} \mathcal{Q}_1(\underline{\mathbf{y}}_{\mathrm{i}})\langle(\mathbf{P}^{T} \otimes \mathbf{I_{N_{\mathrm{r}}}})_\mathrm{i},\underline{\mathbf{G}}(z)\rangle s +  ||\underline{\mathbf{G}}(z)||_*.
\end{equation}
This heuristically designed loss function attempts to minimize the nuclear norm $||\underline{\mathbf{G}}(z)||_*$ of the output of the generator $\mathbf{G}(z)$ while maximizing the correlation between $\mathcal{Q}_1(\mathbf{y})$ (which is $\pm 1$) and $\langle(\mathbf{P}^{T} \otimes \mathbf{I_{N_{\mathrm{r}}}}),\underline{\mathbf{G}}(z)\rangle$. The summation in \eqref{eq:gan_eq_rep_obq} should be interpreted as the sum over the real and imaginary parts, separately,
\begin{equation}
 \sum_{\mathrm{i}=1}^{N_{\mathrm{p}}N_{\mathrm{r}}} \mathcal{Q}_1(\underline{\mathbf{y}}_{\mathrm{i,real}})\langle(\mathbf{P}^{T} \otimes \mathbf{I_{N_{\mathrm{r}}}})_{\mathrm{i,real}},\underline{\mathbf{G}}(z)_{\mathrm{real}}\rangle s + \sum_{\mathrm{i}=1}^{N_{\mathrm{p}}N_{\mathrm{r}}} \mathcal{Q}_1(\underline{\mathbf{y}}_{\mathrm{i,imag}})\langle(\mathbf{P}^{T} \otimes \mathbf{I_{N_{\mathrm{r}}}})_{\mathrm{i,imag}},\underline{\mathbf{G}}(z)_{\mathrm{imag}}\rangle s
\end{equation}

\begin{algorithm} \label{alg:CS_GAN}
\setstretch{1}
\SetAlgoNoLine
1. Train a GAN using a set of realistic channel realizations.\\
2. Extract the trained generator $\mathbf{G}(z)$.\\
3. Given the noisy pilot measurements $\mathbf{\underline{y}}$, reconstruct the channel $\mathbf{\underline{y}}$ encodes by solving \\the following optimization problem using gradient descent:\\
\begin{itemize}
    \item For full resolution pilot measurements:
    \begin{equation*}
        z^* = \underset{z \in \mathbb{R}^d} {\mathrm{arg\ min\ }} \hspace{0.05 in} ||\underline{\mathbf{y}} - (\mathbf{P}^{T} \otimes \mathbf{I_{N_{\mathrm{r}}}})\mathbf{\underline{G}}(z)s||_{2}^2 + \lambda_{\mathrm{reg}} ||z||_{2}^2,
    \end{equation*}
    \item For quantized pilot measurements:
    \begin{equation*} 
        z^* = \underset{z \in \mathbb{R}^d} {\mathrm{arg\ min\ }} \hspace{0.05 in} -\lambda_{\mathrm{reg}} \sum_{\mathrm{i}=1}^{N_{\mathrm{p}}N_{\mathrm{r}}} \mathcal{Q}_1(\underline{\mathbf{y}}_{\mathrm{i}})\langle(\mathbf{P}^{T} \otimes \mathbf{I_{N_{\mathrm{r}}}})_\mathrm{i},\underline{\mathbf{G}}(z)\rangle s +  ||\underline{\mathbf{G}}(z)||_*.
    \end{equation*}
\end{itemize}
\quad The initial point $z_0$ for gradient descent is sampled from $\mathbb{P}_z$.\\
4. The reconstructed channel estimate is then $\mathbf{G}(z^*)$, which is of dimensions $N_{\mathrm{r}} \times N_{\mathrm{t}}$
\caption{Channel Estimation using Deep Generative Networks}
\end{algorithm}

\subsection{Beamforming using the Generative Channel Estimator} 
Having recovered the channel estimate $\mathbf{G}(z^*)$ from the compressed pilot measurements at the UE, we now use this channel estimate to design the optimum RF and baseband precoder $\mathbf{F}_{\mathrm{RF}}$ and $\mathbf{F}_{\mathrm{BB}}$. The optimal latent input vector $z^*$ of the generator $\mathbf{G}$ provides a compressed representation of the channel. If we could convey the weights and architecture of the generator from the UE to the BS during the initial access phase, then in subsequent data transmissions, the CSI overhead would be considerably reduced. At every coherence time, we would simply feedback $z^*$ to the BS and use $\mathbf{G}(z^*)$ as the channel estimate to design the precoder matrices $\mathbf{F}_{\mathrm{RF}}$ and $\mathbf{F}_{\mathrm{BB}}$.

\section{Simulation Details \& Benchmarks} \label{sec:Simulations}

The performance metric is normalized mean square error (NMSE), defined as 
\begin{equation} \label{eq:NMSE}
\text{NMSE} = \mathbb{E}\left[\frac{||\underline{\mathbf{H}}-\underline{\hat{\mathbf{H}}}||_2^2}{||\underline{\mathbf{H}}||_2^2}\right],
\end{equation}
where $\underline{\mathbf{H}}$ and $\underline{\hat{\mathbf{H}}}$ are column vectors that specify the actual and estimated channel taps in the frequency domain over all antennas, respectively.

\subsection{Data Generation}
Channel realizations have been generated using the 5G Toolbox in MATLAB in accordance with the 3GPP specifications TR 38.901\footnote{https://www.etsi.org/deliver/etsi\_ tr/138900\_138999/138901/14.00.00\_60/tr\_138901v140000p.pdf}. 
The channel simulation parameters are listed in Table \ref{tab:channel_param}. In order to generate structure in the channel realizations, some degree of correlation is required between neighbouring antennas at the BS and the UE. To generate this correlation, the antenna element spacing in the uniform linear arrays (ULA) at the BS and UE were assumed to be $\lambda/10$. This reduced antenna spacing is a crucial assumption, and we will justify its requirement in Section \ref{subsec:explanations}. Each channel realization generated in MATLAB was of dimension $(N_{\mathrm{f}},12,N_{\mathrm{r}},N_{\mathrm{t}})$, the first and second dimension being the number of subcarriers and number of OFDM symbols respectively. To focus on exploitation of the spatial structure of the channel matrices, we simply extract the $(N_{\mathrm{r}},N_{\mathrm{t}})$ matrix corresponding to the first subcarrier and first OFDM symbol for the purpose of these simulations.  
\begin{table}
	\caption[Simulation Parameters] {Simulation Parameters}
	\label{tab:channel_param}
	\begin{tabular}{ |p{6cm}|p{4cm}|}
		\hline
		Delay Profile & CDL-D \\
		\hline
		Subcarrier Spacing & 15 kHz\\
		\hline
		$N_{\mathrm{t}}$ & 64 \\
		\hline
		$N_{\mathrm{r}}$ & 16 \\
		\hline
        Antenna Array Type & ULA\\
 		\hline
 		Antenna Spacing & $\lambda/10$\\
 		\hline
		Sampling Rate & 15.36 MHz\\
		\hline
		Carrier Frequency & 40 GHz\\
		\hline
		Delay Spread & 30 ns \\
		\hline
		Doppler Shift & 5 Hz\\
		\hline
		 $N_{\mathrm{f}}$ & 14\\
		\hline
	\end{tabular}\centering
\end{table}

\subsection{Data Pre-processing}
Note that $\mathbf{G}(z)$ has dimensions $(N_{\mathrm{t}},N_{\mathrm{r}},2)$, where the last entry corresponds to the real and imaginary part. Thus, in the training dataset, $\mathbf{H}$ has to be split up into its real and imaginary part and concatenated to obtain $\mathbf{H}_\mathbf{G} \in \mathbb{R}^{N_{\mathrm{r}} \times N_{\mathrm{t}} \times 2}$, while $\mathbf{G}(z)$ has to be reshaped as a complex-valued matrix before being utilized for optimization in \eqref{eq:gan_eq_rep} or \eqref{eq:gan_eq_rep_obq}. Before using the data for training the GAN, we normalize the channel matrices $\mathbf{H}_\mathbf{G} \in \mathbb{R}^{N_{\mathrm{r}} \times N_{\mathrm{t}} \times 2}$ element-wise as \begin{equation} \label{eq:normalization}
    \mu_{i} = \mathbb{E}[\mathbf{H}_{\mathbf{G}i}] \; \; \;
    \sigma_{i}^2 = \mathbb{E}[(\mathbf{H}_{\mathbf{G}i} - \mu_{i})^2]
\end{equation}
\begin{equation} \label{eq:normalization_apply}
    \mathbf{H}_{\mathbf{G}i,norm} = \frac{\mathbf{H}_{\mathbf{G}i} - \mu_{i}}{\sigma_{i}},
\end{equation}
where $i \in [2N_{\mathrm{t}}N_{\mathrm{r}}]$ and subscript $i$ is used to denote the $i^{th}$ element in the array. While testing, we do not have access to the element-wise mean and variance, hence we continue to use the training mean and variance. This implies that $\mathbf{G}(z)$ in \eqref{eq:gan_eq_rep} is replaced by \begin{equation} \label{eq:gan_replace}
    \mathbf{G}(z)_{i} \leftarrow \mu_{i} + \sigma_{i}\mathbf{G}(z)_{i}  .
\end{equation} We performed a simulation to ascertain the impact of this artifact, and found it was negligible. The need for normalization arises from empirical evidence that the GAN is unable to learn mean-shifted distributions \cite{srivastava2017veegan}. 
\subsection{NN Architecture of Generator}
The GAN was implemented in Keras\footnote{https://github.com/fchollet/keras} and PyTorch\footnote{https://github.com/pytorch/pytorch}, with the basic implementation given online\footnote{https://github.com/eriklindernoren/Keras-GAN}. The generator and discriminator employed in the Wasserstein GAN were Deep Convolutional NNs. While the discriminator architecture was adopted from \cite{arjovsky2017wasserstein}, the generator was fine-tuned to improve its ability to learn the underlying probability distribution
and its architecture is described next.

The generator $\mathbf{G}$ takes an input $z \in \mathbb{R}^d$, passes it through a dense layer with output size $128N_{\mathrm{t}}N_{\mathrm{r}}/16$, and reshapes it to an output size of $(N_{\mathrm{t}}/4,N_{\mathrm{r}}/4,128)$. This latent representation is then passed through $k=2$ layers, each consisting of the following units: upsampling, 2D Convolution with a kernel size of 4 and Batch Normalization. At each stage, $2 \times 2$ upsampling is performed, i.e. the input is reshaped from $(N_{\mathrm{t}}/n,N_{\mathrm{r}}/n,128)$ to $(2N_{\mathrm{t}}/n,2N_{\mathrm{r}}/n,128)$ by replicating the corresponding values. The performance of the generator is sensitive to this choice of sampling factor, with oversampling of 4 and above preventing the generator from learning the channel distribution. Similarly, a kernel size of 4 corresponds to using a $4 \times 4$ filter in the first two dimensions to replace each value by a weighted average of the neighboring values that are within a $4 \times 4$ square surrounding it. Both upsampling and 2D convolution thus model the local correlations in a spatial channel matrix, with larger upsampling and size of kernel filter corresponding to a greater estimated spatial correlation. It is finally passed through a 2D Convolutional layer with a kernel size of 4 and linear activation to obtain $\mathbf{G}(z)$, the $N_{\mathrm{r}} \times N_{\mathrm{t}}$ channel estimate.

\subsection{GAN Training Details \& GCE}
The training and test parameters for the Wasserstein GAN are specified in Table \ref{tab:training_param}. The generator thus obtained is utilized in the GCE, to find the optimal $z^*$ for each channel realization in the test dataset. To minimize the loss function in \eqref{eq:gan_eq_rep} or \eqref{eq:gan_eq_rep_obq}, as the case may be, we use two approaches. A derivative-free optimization procedure known as Powell's conjugate direction method \cite{powell1964efficient}, with a relative error tolerance of $\epsilon = 10^{-5}$ was employed in minimizing \eqref{eq:gan_eq_rep} and \eqref{eq:gan_eq_rep_obq} for the generative model trained in Keras, since a trained Keras model does not provide for differentiation of the loss function in \eqref{eq:gan_eq_rep} with respect to the input vector $z$. However, as explained in \cite{paszke2017automatic}, PyTorch allows automatic differentiation and hence an Adam \cite{kingma2014adam} optimizer with a learning rate of $\eta = 10^{-2}$ and iteration count of 100 is utilized in minimizing \eqref{eq:gan_eq_rep}, for the generative model trained in PyTorch.

\begin{table}
	\caption{GAN Training Parameters}
	\label{tab:training_param}
	\begin{tabular}{ |p{6cm}|p{4cm}|}
	    \hline
	    Training dataset size & 3654\\
	    \hline
	    Test dataset size & 12 \\
	    \hline
	    Optimizer & RMSProp\protect\footnotemark\\
	    \hline 
	    Learning Rate & 0.00005\\
	    \hline
	    Batch size & 200\\
	    \hline
	    Epochs & 3000\\
	    \hline
	    $\lambda_{\mathrm{reg}}$ & 0.001\\
	    \hline
	\end{tabular}\centering
\end{table} \footnotetext{https://www.cs.toronto.edu/~tijmen/csc321/slides/lecture\_slides\_lec6.pdf} 
\subsection{Compressed Sensing Based DL Channel Estimation} \label{subsec:baselines}
In this subsection, we describe the baselines used for assessing the performance of GCE. Since we consider the narrow-band clustered channel model, we can use the virtual channel model \cite{sayeed2002deconstructing} to obtain a sparse representation of the channel matrix in the DFT basis. More specifically, assuming uniform spaced linear arrays at the transmitter and receiver, the array response matrices are given by the unitary DFT matrices $\mathbf{A}_{\mathrm{T}} \in \mathbb{C}^{N_{\mathrm{t}} \times N_{\mathrm{t}}}$ and $\mathbf{A}_{\mathrm{R}} \in \mathbb{C}^{N_{\mathrm{r}} \times N_{\mathrm{r}}}$. Then we can represent $\mathbf{H}$ in terms of a $\mathrm{K}$-sparse matrix $\mathbf{H}_\mathrm{v} \in \mathbb{C}^{N_{\mathrm{r}} \times N_{\mathrm{t}}}$
\begin{equation} \label{eq:virtual_channel_rep}
\begin{aligned}
    &\mathbf{H} = \mathbf{A}_{\mathrm{R}}\mathbf{H}_\mathrm{v}\mathbf{A}_{\mathrm{T}}^H \\
    &\underline{\mathbf{H}} = ((\mathbf{A}_{\mathrm{T}}^H)^T \otimes \mathbf{A}_{\mathrm{R}})\underline{\mathbf{H}_\mathrm{v}}.
\end{aligned}
\end{equation}
Therefore, the received signal at the UE $\underline{\mathbf{y}}$ is given by \begin{equation}
    \underline{\mathbf{y}} = ((\mathbf{A}_{\mathrm{T}}^HP)^T \otimes \mathbf{A}_{\mathrm{R}})\underline{\mathbf{H}_\mathrm{v}} s + \underline{\mathbf{n}}.
\end{equation}
Denoting by $\mathbf{A}_{\mathrm{sp}} = ((\mathbf{A}_{\mathrm{T}}^HP)^T \otimes \mathbf{A}_{\mathrm{R}})$, as explained in \cite{mendez2016hybrid}, the reconstruction of the channel can be formulated as a non-convex combinatorial problem \begin{equation} \label{eq:sparse_min}
    \underset{\underline{\mathbf{H}_\mathrm{v}} \in \mathbb{C}^{N_{\mathrm{r}}N_{\mathrm{t}}}}{\textrm{minimize } } ||\underline{\mathbf{H}_\mathrm{v}}||_0 \textrm{ subject to } ||\underline{\mathbf{y}} - \mathbf{A}_{\mathrm{sp}}\underline{\mathbf{H}_\mathrm{v}}s||_2 \leq \sigma.
\end{equation}
A variety of Matching Pursuit (MP) and Approximate Message Passing (AMP) algorithms have been proposed to solve \eqref{eq:sparse_min}. In particular, we consider three approaches:

i)\textbf{ Orthogonal Matching Pursuit (OMP):} We directly solve \eqref{eq:sparse_min} using OMP, as described in \cite{mendez2016hybrid}. The stopping criterion for OMP is based on the power of the residual error. We stop when the energy in the residual is smaller than a given threshold \footnote{The maximum number of iterations are set to be 100. If OMP is allowed to run further, it fits to the noise at low SNR and the NMSE increases.}, which is chosen to be $\sigma^2$.

ii) \textbf{Lasso Baseline:} Consider the $\mathcal{L}_1$ convex relaxation of \eqref{eq:sparse_min}, and use Basis Pursuit Denoising to solve this problem. In its Lagrangian form, it can be written as: 
\begin{equation}
    \underset{\underline{\mathbf{H}_\mathrm{v}} \in \mathbb{C}^{N_{\mathrm{r}}N_{\mathrm{t}}}}{\textrm{minimize } } ||\underline{\mathbf{H}_\mathrm{v}}||_1  + \lambda_{sp}||\underline{\mathbf{y}} - \mathbf{A}_{\mathrm{sp}}\underline{\mathbf{H}_\mathrm{v}}s||_2.
\end{equation}
However all the norms and matrices involved are complex valued. Hence, an $\mathcal{L}_1$ norm minimization problem gets converted into a second order conic programming (SOCP) problem \cite{winter2005real}, and can be solved by standard convex solvers such as CVXPY \cite{cvxpy}.

iii) \textbf{EM-GM-AMP:} Approximate Message Passing algorithms such as EM-GM-AMP \cite{vila2013expectation} are well-established Bayesian techniques for sparse signal recovery from noisy compressive linear measurements that are known to hold for a large class of sensing matrices. Using the EM-GM-AMP implementation described in \cite{vila2013expectation}, 
we input $\underline{\mathbf{y}}$ and $\mathbf{A}_{\mathrm{sp}}$ and recover the channel estimate $\underline{\mathbf{H}_\mathrm{v}}$, which is then used to recover $\mathbf{H}$ using the array response matrices $\mathbf{A}_{\mathrm{T}}$ and $\mathbf{A}_{\mathrm{R}}$. It is to be noted that assuming an antenna spacing of $\lambda/10$, with the columns of $\mathbf{A}_{\mathrm{T}}$ as well as $\mathbf{A}_{\mathrm{R}}$ being independent, leads to the entries of $\underline{\mathbf{H}_\mathrm{v}}$ being correlated. This correlation is however not exploited by EM-GM-AMP. Improved benchmarking comparisons with algorithms such as EMturboGAMP \cite{schniter2010turbo} that attempt to exploit structured sparsity in non i.i.d signals is left for future work.

These are the three sparse signal recovery baselines - that each require knowledge of the sparsifying basis - we use to assess the performance of the proposed GCE. It should be noted that the beamspace sparsity that is exploited by the CS algorithms is in no way utilized by GCE.

\section{Results} \label{sec:results}

\begin{figure}[!ht]
    \centering
    \includegraphics[width=5in]{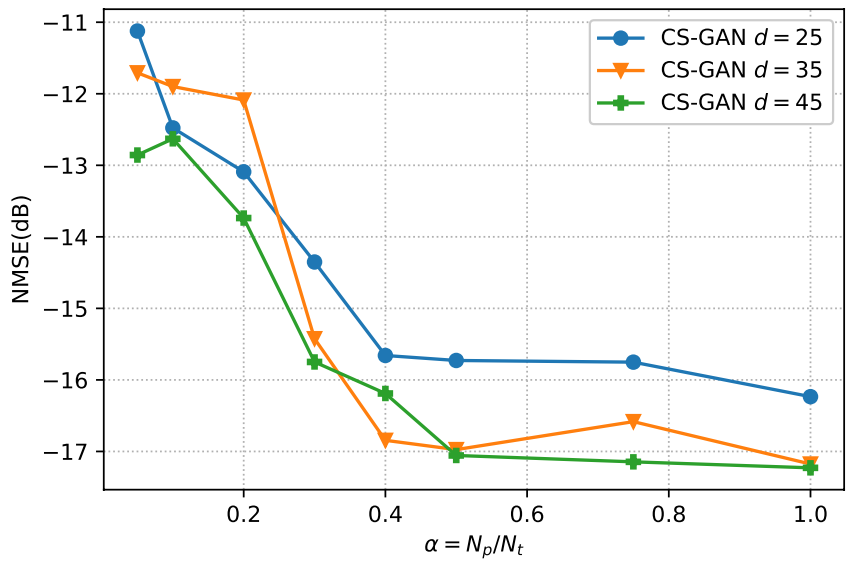}
    \caption{NMSE vs. $\alpha = N_{\mathrm{p}}/N_{\mathrm{t}}$ for varying dimension $d$ of the input $z$ to the generator $\mathbf{G}$}
    \label{fig:nmse_alpha}
\end{figure}

The first experiment performed is to determine the optimal latent dimension $d$ of the input $z$ to the generator. Ideally, CS techniques would determine $d$ in the absence of noise, hence we fix the SNR at a high value of 40 dB, and evaluate the NMSE as a function of the number of pilot measurements $N_{\mathrm{p}}$ for varying values of $d$ as shown in Fig \ref{fig:nmse_alpha} using full resolution measurements. 

From Fig \ref{fig:nmse_alpha}, we can see that $d=35$ appears sufficient with $N_{\mathrm{p}}/N_{\mathrm{t}} = 0.4$.
Increasing the number of pilot measurements $N_{\mathrm{p}}/N_{\mathrm{t}}$ beyond 0.4 does not have any measurable impact on the NMSE. This indicates that any more measurements would not improve the accuracy of the channel prediction. More importantly, it highlights that there exists a compressed representation for the channel in an unknown basis, but using the optimal latent input vector $z^*$ defined in \eqref{eq:gan_eq_rep}, we can recover the channel prediction perfectly without knowing, for example, that the channel is sparse in the DFT basis. We obtain a nearly 50x compressed representation of the channel,  with under 40 parameters needed to represent a $16 \times 64$ channel matrix realization( = 2048 real values). While current mmWave channel estimation techniques focus on the optimal design of training precoders and combiners under the assumption of either virtual channel models \cite{sayeed2002deconstructing}, UIU models \cite{tulino2006capacity} among others, the GCE minimizes the need for their optimal design and provides a model-free approach for representing inherently sparse or structured channels. This may prove valuable for future deployments at progressively higher carrier frequencies, where these models may not hold.  
With $d = 35$, we now vary the SNR, and observe the NMSE vs. SNR for varying $\alpha = N_{\mathrm{p}}/N_{\mathrm{t}}$ in the case of full-resolution and one-bit quantized pilot measurements. The OMP, Lasso and EM-GM-AMP baselines are also plotted.

\subsection{Full Resolution Channel Estimation}
\begin{figure}[!ht]
    \centering
    {\includegraphics[height=0.25\textheight, width=0.47\textwidth]{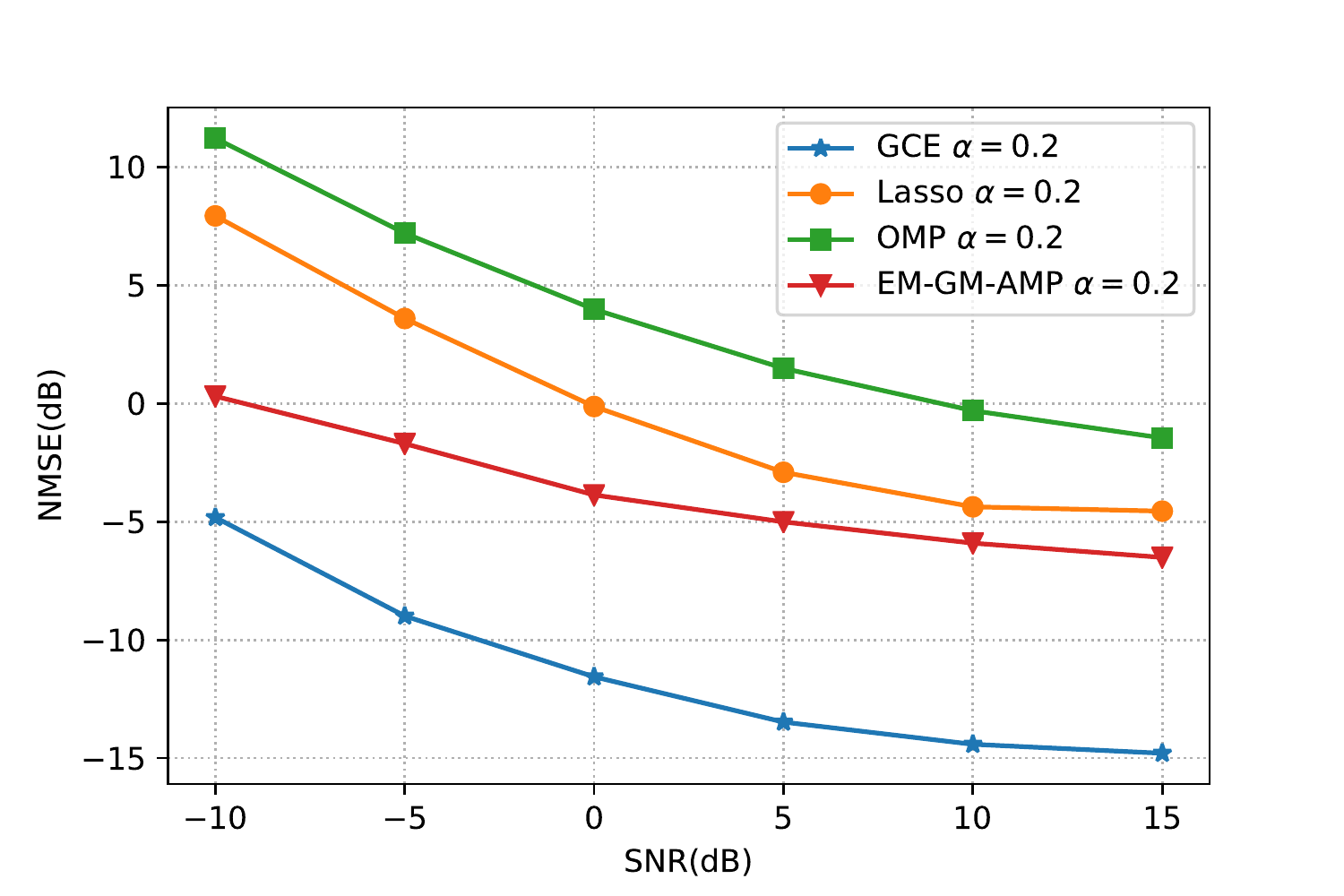}
    \label{fig:nmse_snr_alpha_20}}
    \hspace{0.2in}
    {\includegraphics[height=0.25\textheight, width=0.47\textwidth]{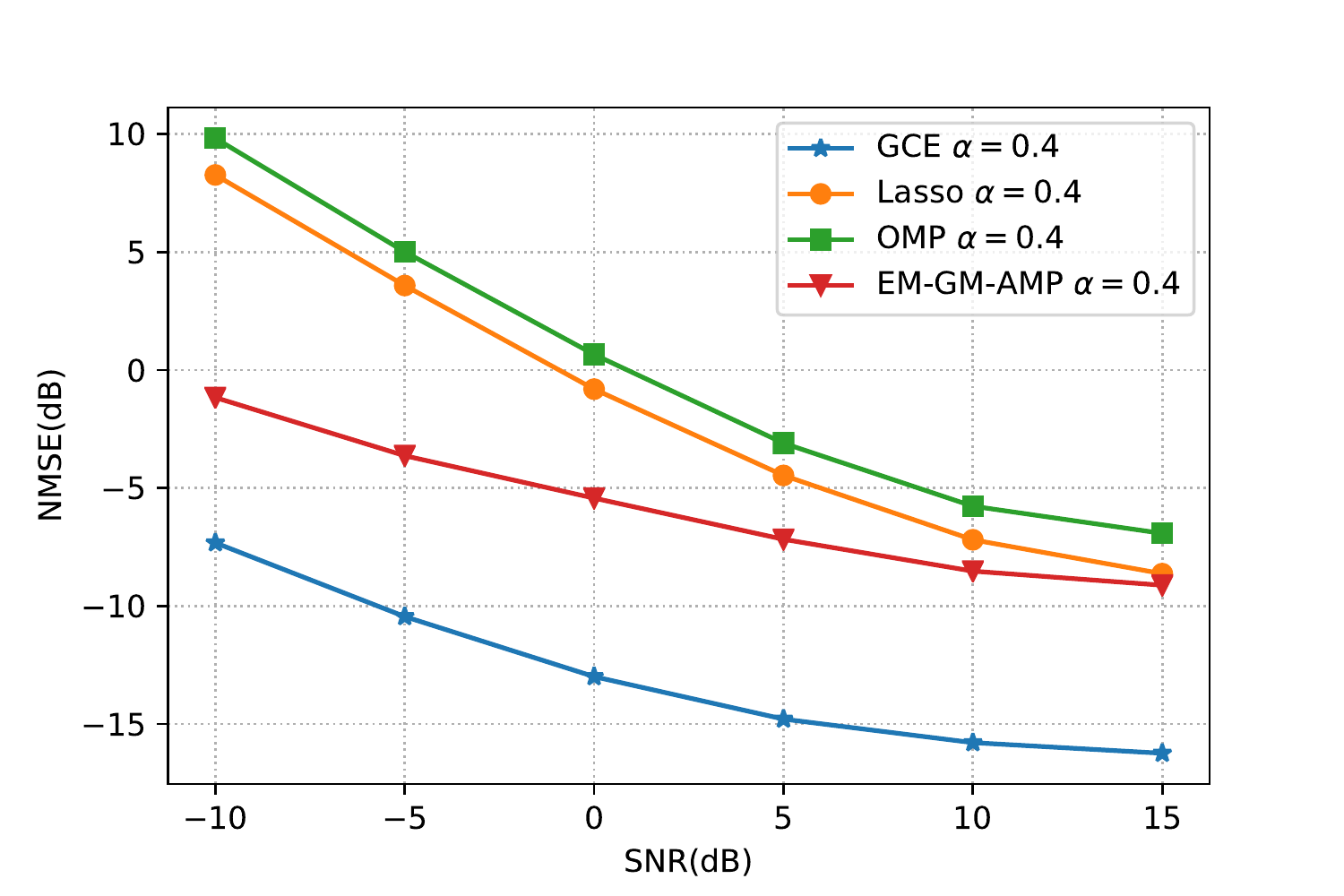}\label{fig:nmse_snr_alpha_40}}\\
    {\includegraphics[height=0.25\textheight, width=0.47\textwidth]{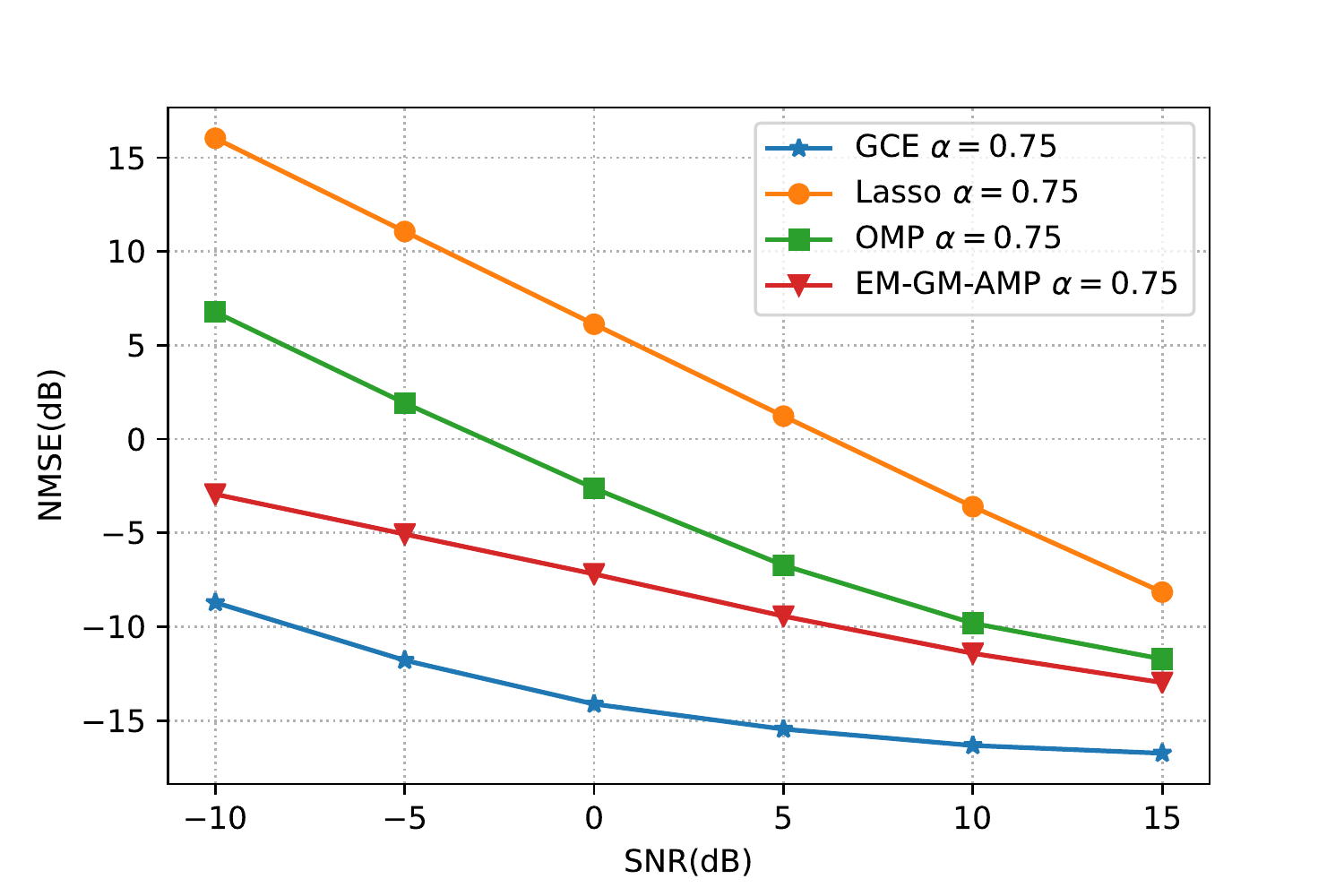}\label{fig:nmse_snr_alpha_75}}
    \hspace{0.2in}
    {\includegraphics[height=0.25\textheight, width=0.47\textwidth]{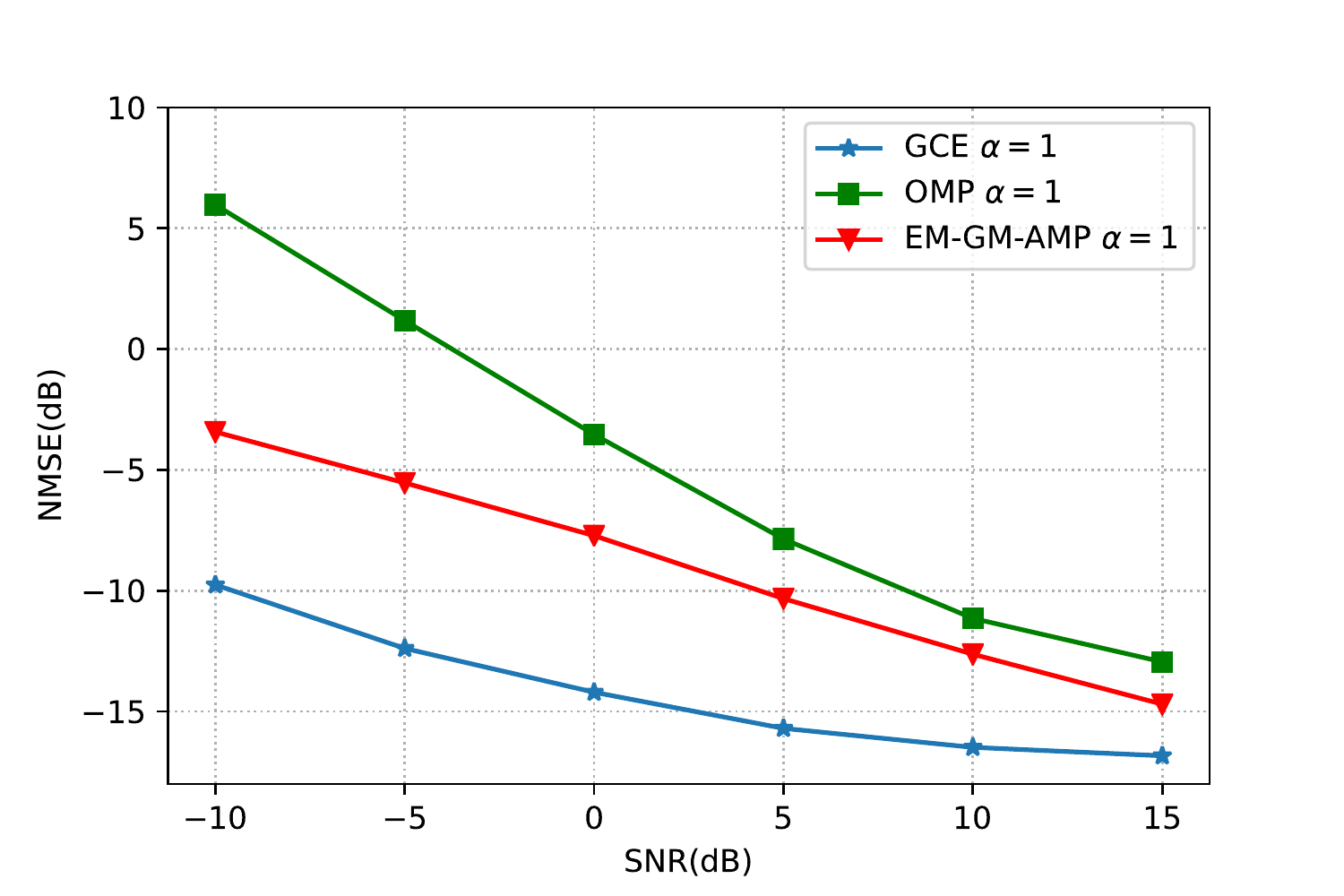}\label{fig:nmse_snr_alpha_100}}
    \caption{NMSE vs. SNR for various values of $\alpha = N_{\mathrm{p}}/N_{\mathrm{t}}$. The $\alpha$ values are [0.2, 0.4, 0.75, 1]. The Lasso curve is omitted for $\alpha = 1$ since CVXPY \cite{cvxpy} takes too long to converge due to the large number of optimization variables.}
    \label{fig:nmse_snr}
\end{figure}

 As shown in Fig \ref{fig:nmse_snr}, the GCE offers large improvement in NMSE, of at least 5 dB at an SNR of -10 dB and up to 8 dB at an SNR of 15 dB for $\alpha = 0.2$ over the EM-GM-AMP baseline. The GCE's performance also does not change significantly as $\alpha$ increases from 0.4 to 0.75, indicating that the prior learnt by the generator $\mathbf{G}$ is informative enough to require only $40\%$ of the total number of pilots that would have been needed by a well-posed channel estimation problem to reconstruct the channel. Moreover, the improvement in NMSE offered by the GCE decreases as $\alpha$ increases from 0.2 to 1, with the gap between EM-GM-AMP and GCE being reduced to 2 dB at an SNR of 15 dB and $\alpha = 0.1$. However, at low and medium SNR, the GCE continues to outperform all CS based methods significantly. 
 
\subsection{One-bit Quantized Channel Estimation} \label{subsec:obq}
\begin{figure}[!ht]
    \centering
    \includegraphics[width=5in]{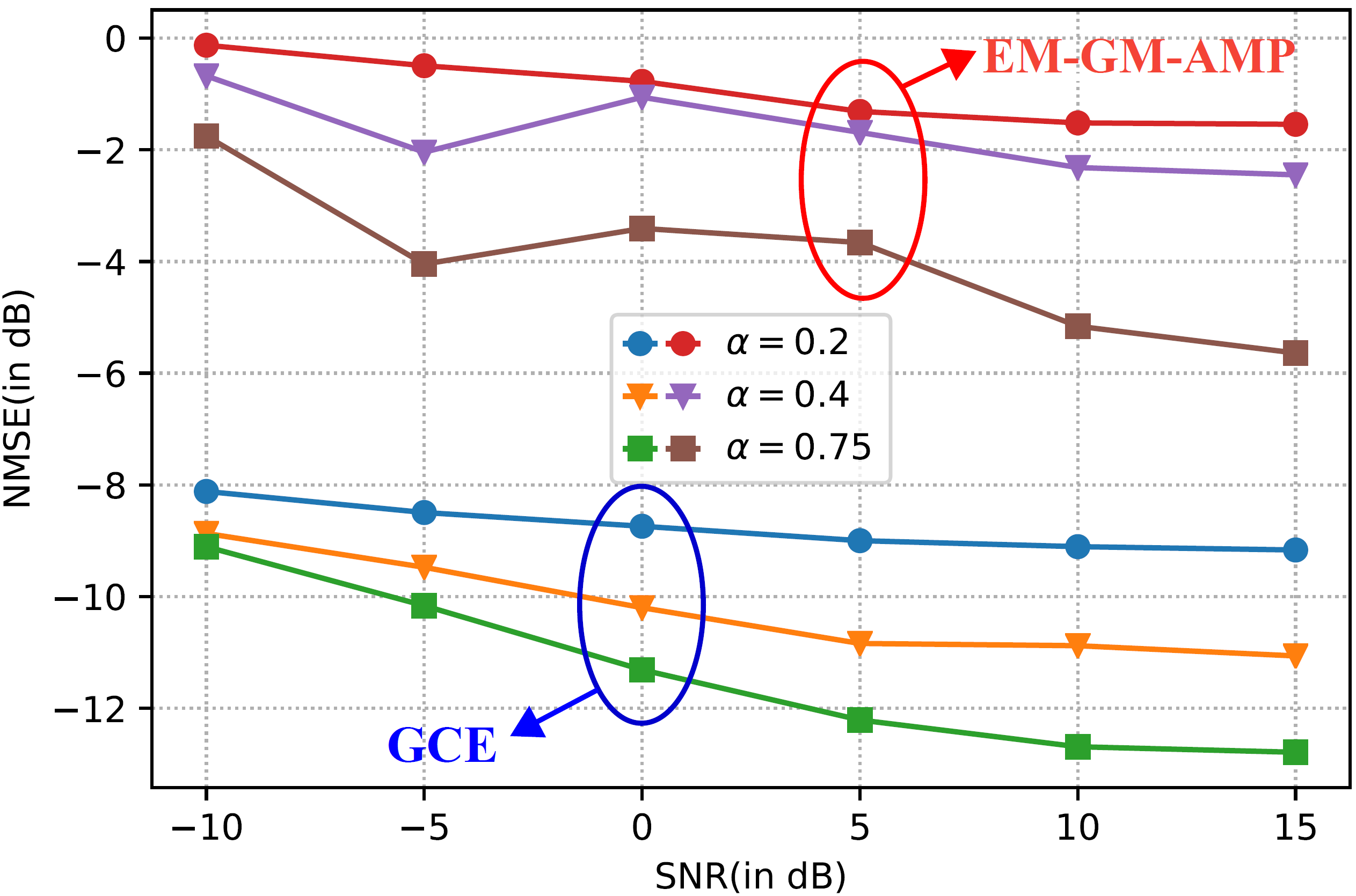}
\caption{NMSE v/s SNR as a function of $ \alpha = N_{\mathrm{p}}/N_{\mathrm{t}}$ with one-bit quantization.}
    \label{fig:nmse_obq}
\end{figure}

The NMSE for the case of 1-bit quantized pilot measurements is defined slightly differently, since in one-bit measurements, we cannot determine the relative scaling factor for the reconstructed channel matrices.
\begin{equation}
    \text{NMSE} = \mathbb{E}\left[\frac{||\underline{\mathbf{H}}- \kappa\underline{\hat{\mathbf{H}}}||_2^2}{||\underline{\mathbf{H}}||_2^2}\right],
\end{equation}
where $\kappa = \mathrm{argmin}||\underline{\mathbf{H}}- \kappa\underline{\hat{\mathbf{H}}}||_2^2$ for a given $\underline{\mathbf{H}}$ and $\underline{\hat{\mathbf{H}}}$. Note that though this may seem genie-aided, precoder optimization that finally determines the achievable rate is not affected by this scaling factor. The dependence of NMSE on SNR for one-bit measurements with varying number of pilots is shown in Fig. \ref{fig:nmse_obq}, and contrasted with the performance of EM-GM-AMP on the same measurements. As one can clearly see, the GCE brings about an immense improvement in NMSE, and this can be attributed to the rich prior stored in the weights of the generator.

\subsection{Hybrid Precoding for Quantized Channel Estimation}
\begin{figure}[!ht]
    \centering
    \includegraphics[width=5in]{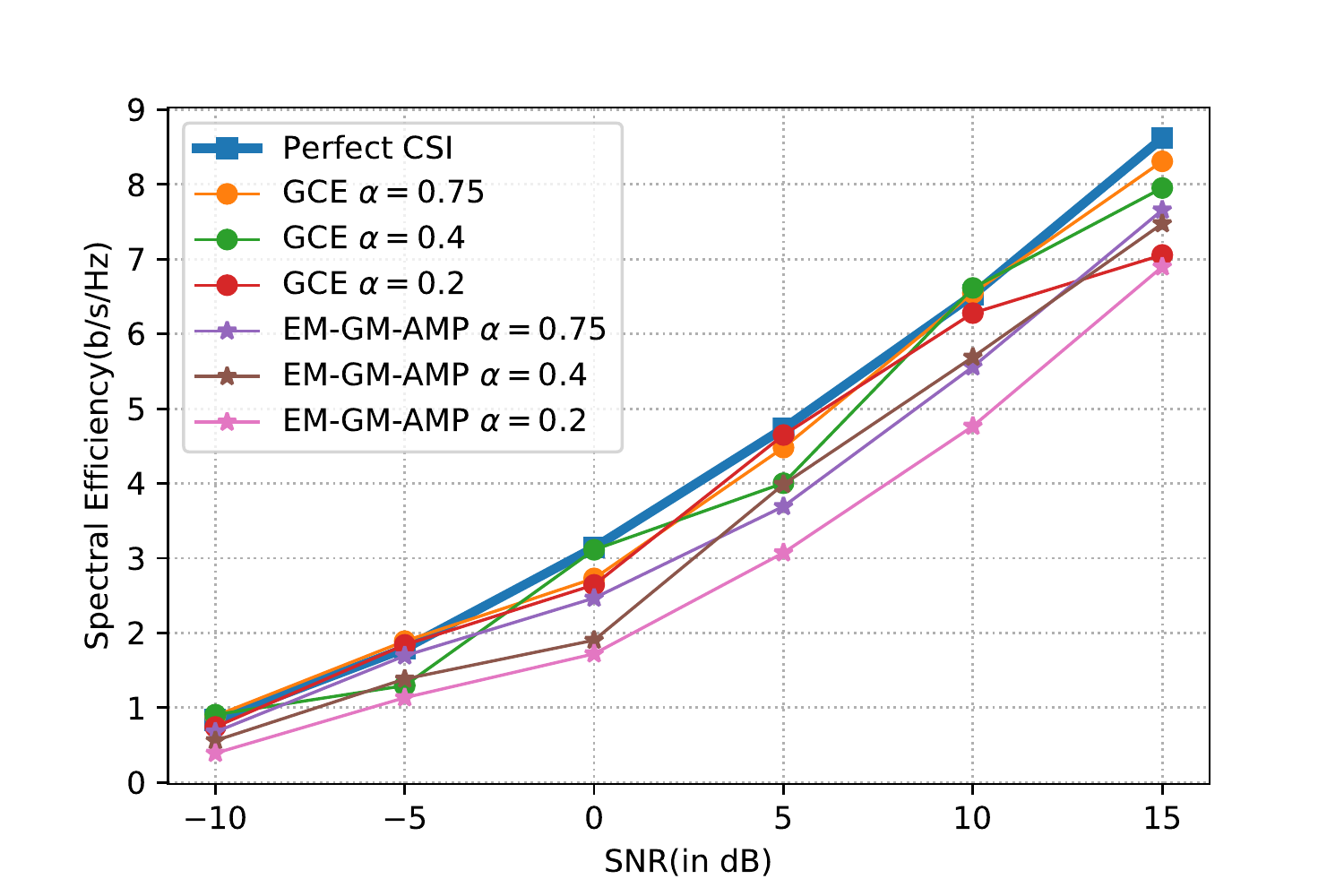}
\caption{Spectral Efficiency v/s SNR as a function of $ \alpha = N_{\mathrm{p}}/N_{\mathrm{t}}$ with one-bit quantization using OMP-based precoding}
    \label{fig:nmse_obq_cap}
\end{figure}

To validate the improvement in channel estimate quality postulated in Section \ref{subsec:obq}, we calculate the spectral efficiency obtained using hybrid precoding in the data transmission phase. We assume $N_{\mathrm{s}} = \mathrm{min}(N_{\mathrm{t}},N_{\mathrm{r}}) = 16$ and optimal unconstrained combiners are employed at the UE. The RF and baseband precoders $\mathbf{F}_{\mathrm{RF}}$ and $\mathbf{F}_{\mathrm{BB}}$ are computed as explained in Section \ref{sec:system_model} using OMP. Three different channel estimates are used for designing these precoders: the estimate returned by the GCE, the AMP algorithm EM-GM-AMP \cite{vila2013expectation} and the ground truth channel realization (for computing the perfect CSI curve). The spectral efficiency vs. SNR plots are shown in Fig. \ref{fig:nmse_obq_cap} for varying $\alpha$. As is evident, the GCE channel estimate enables design of precoders that support higher capacity data transmissions than EM-GM-AMP.

\subsection{Explanations \& Caveats} \label{subsec:explanations}
\begin{figure}
    \centering
    \subfigure[Singular values of channel realizations in descending order of magnitude]{
        \includegraphics[height=0.20\textheight, width=0.45\textwidth]{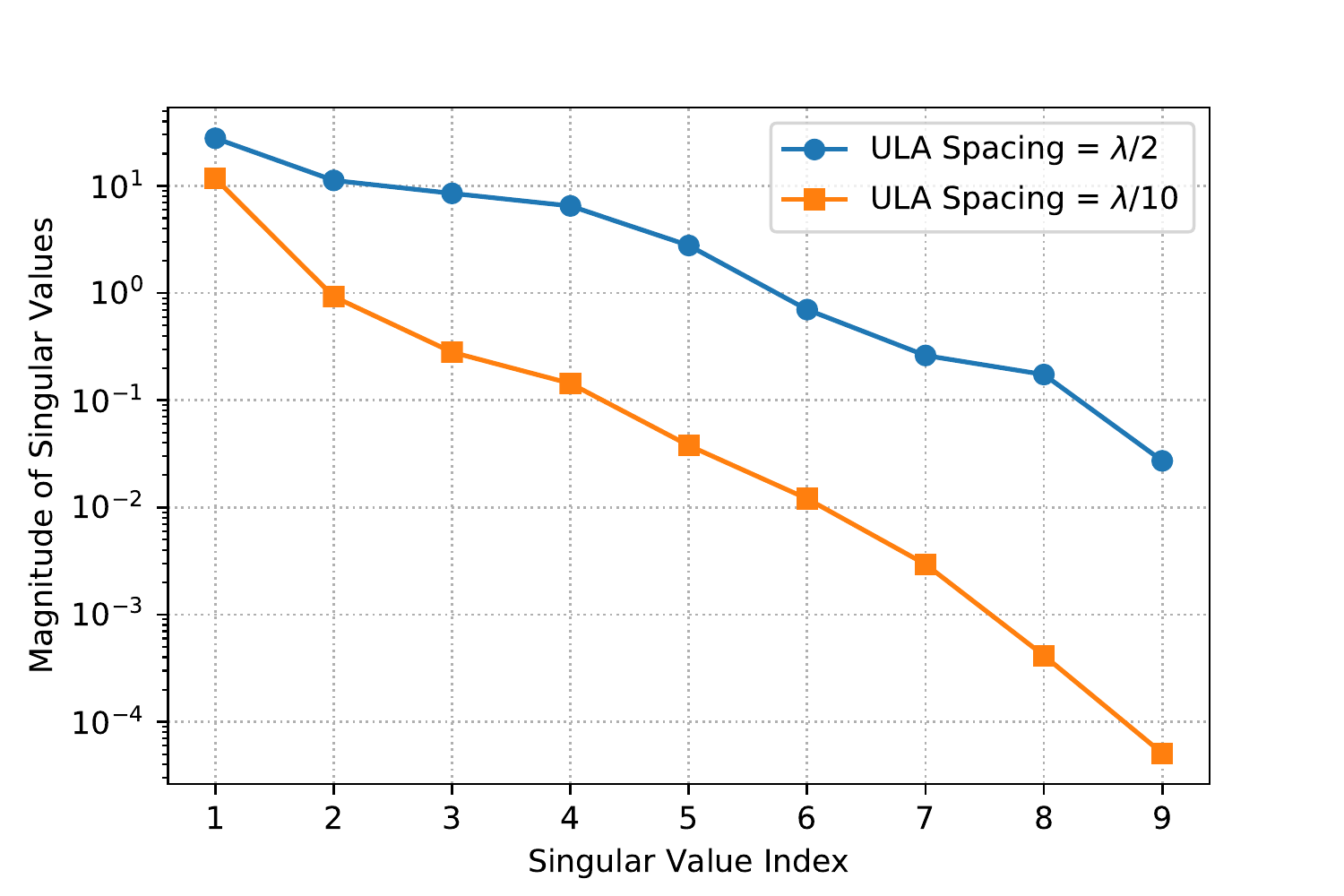}
        \label{subfig:sing_values}}
    \hspace{0.07in}
    \subfigure[NMSE v/s SNR for the two datasets of channel realizations with antenna spacing $\lambda/2$ and $\lambda/10$.]{
        \includegraphics[height=0.20\textheight, width=0.45\textwidth]{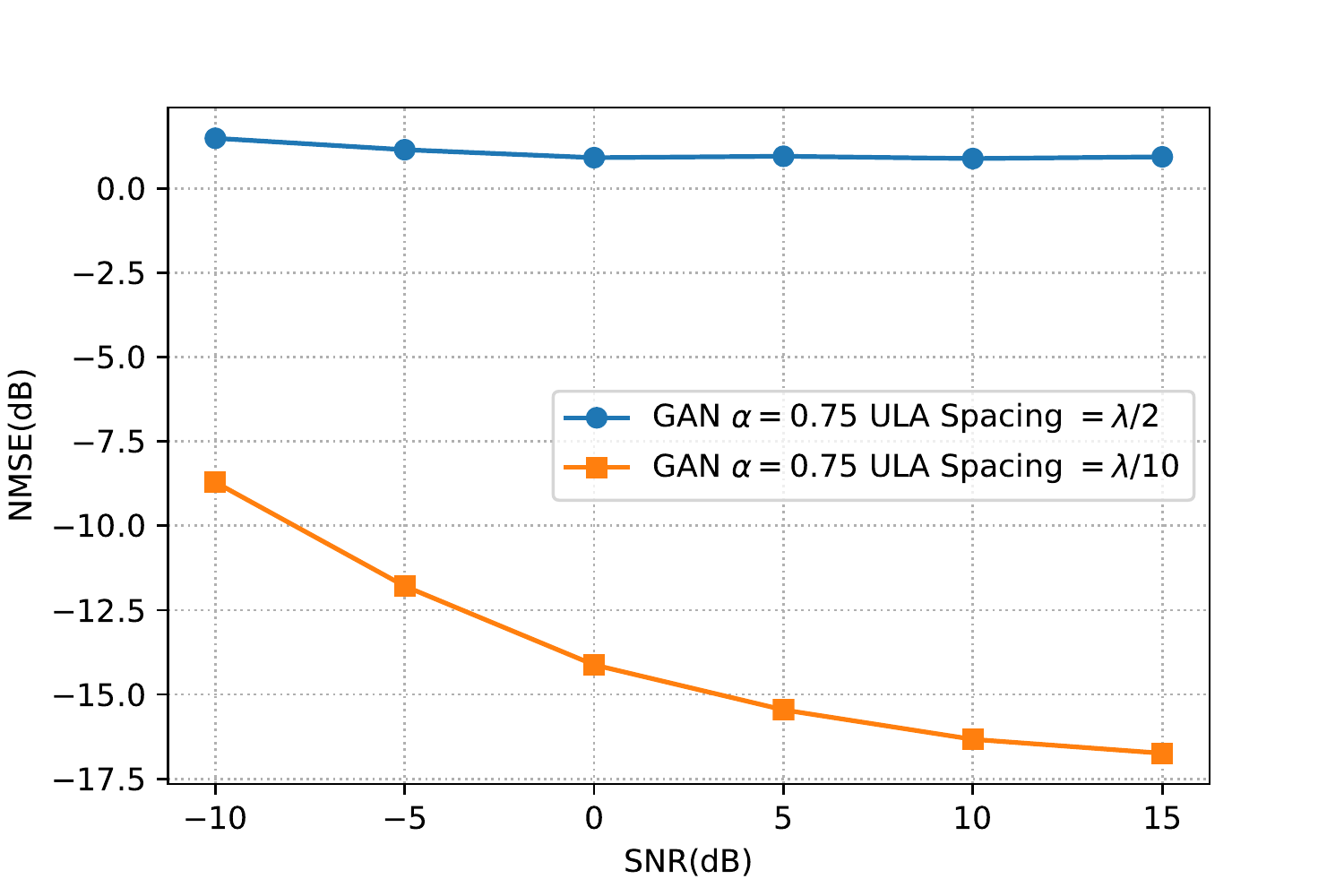}
        \label{subfig:nmse_lambda}}
    \caption{The left figure shows the singular value profile of a channel realization with an antenna spacing of $\lambda/10$ and $\lambda/2$. The higher correlation in the $\lambda/10$ realization enables the generator to learn a rich prior and the GCE to obtain a significantly lower NMSE as shown in the figure on the right.}
    \label{fig:sing_values_comp}
\end{figure}

The benefit obtained from the GCE is clear in the low pilot density and low SNR regime. As the number of pilot symbols increases, the performance of standard CS-based methods gets closer to the GCE, and would be similar to that of the GCE for $N_\mathrm{p} \geq N_\mathrm{t}$. At low SNR, the pilot measurements received are of very poor quality, hence CS-based methods do not perform well, but the GCE utilizes its prior to obtain performance that cannot be achieved by the CS-based methods. This is clearly evident in the one-bit quantized case (Fig. \ref{fig:nmse_obq}), where the GCE curves are roughly parallel to the EM-GM-AMP curves with the constant gap being the generative prior gain.
It can be expected that as the number of antennas packed onto a planar array increases with the move toward THz carrier frequencies, sending an adequate number of pilots would lead to an unsustainable overhead, and recovering the channel estimate from an insufficient number of pilots will become critical. While the GCE outperforms the three CS-based methods, it is important to note the following caveats:

\textbf{High Spatial Correlation:} GCE required a reduced antenna spacing of $\lambda/10$, rather than $\lambda/2$, to successfully learn the channel distribution. As shown in Fig. \ref{subfig:sing_values}, the singular value profile of a $\lambda/2$ channel realization has a higher effective rank than a $\lambda/10$ realization, due to its lower spatial correlation. As a consequence, the generator of a GAN trained on $\lambda/2$ channel realizations was unable to learn the underlying probability distribution and the resulting performance of GCE was poor as shown on the right in Fig. \ref{subfig:nmse_lambda} for $\alpha = 0.75$. Since a GAN was originally designed to learn the prior for image datasets, which have extremely high spatial correlation, the GCE was also found to work in a similar domain. However, as thousands of antennas get deployed at a transmitter or receiver due to their tiny size, it is expected that such singular value profiles will become more commonly observed and only the maximum eigenvector will be needed to acheive capacity in this regime.

A recent paper \cite{hasan2018dual} shows how metamaterial antennas can be used for wireless communications, including LTE and WiFi. Conventional antennas that are very small compared to the wavelength reflect most of the signal back to the source. However, a metamaterial antenna steps-up the antenna's radiated power and behaves as if it were much larger than its actual size, because its novel structure stores and re-radiates energy, which could lead to the deployment of sub-wavelength antennas.

\textbf{Rich Generative Prior:} The weights $\theta_g$ of the generator $G(z;\theta_g)$ encode a probability distribution over the space of permissible spatial channel matrices, such that by inputting $z$, we can draw samples from that distribution. Conventional CS techniques have no such prior knowledge of the distribution of the channel matrices, however they capitalize on the sparsity of the beamspace representation of the channel, which the GCE does not utilize. The results seem to indicate that the generative prior is much more informative than the sparsifying basis, but we have no means of quantifying this yet. Recent efforts in theoretical machine learning \cite{achille2019information} have attempted to quantify the information in the weights of a NN in terms of the impact that perturbing a weight has on the cross-entropy loss. Such work could prove very useful in quantifying the information gain of a generative prior.

\textbf{Training on Simulated Channel Realizations:} We have currently trained a GAN using simulated channel realizations, since obtaining realistic channel data has not proven possible, even with our industry partners. One can only hope to recover the channel estimate based on pilot measurements from current transceiver chips. It remains to be seen if the GAN can succeed in learning the channel distribution even from these noisy channel estimates. The original GAN proposed in \cite{goodfellow2014generative} is known to learn discriminators with poor generalization capabilities, and many recent works \cite{srivastava2017veegan,arjovsky2017wasserstein,thanh2019improving} have taken different approaches to justifying design of custom objective functions for the discriminator that would help the generator to better approximate the target distribution, and improve the generalization capability of the discriminator.

\subsection{Timing Analysis}
Using the PyTorch based generative model, optimization of \eqref{eq:gan_eq_rep} involves only performing gradient descent with respect to $z \in \mathbb{R}^d$, with $d = 35$ in our case. Hence one would expect each iteration to be computationally inexpensive. To determine the computational advantage of using GCE, we perform a comparision of its execution time per iteration as compared to the CS baselines and the results are tabulated in Table \ref{tab:execution_time}. The number of iterations required to achieve the NMSE results in Fig. \ref{fig:nmse_snr} for each method are also given in Table \ref{tab:execution_time}. The evaluation of the first three methods was performed on an Intel i9-8950HK CPU @ 2.90GHz. The results for GCE are given both when performed on the Intel i9-8950HK CPU without acceleration as well as when accelerated using a Nvidia GeForce GTX 2070 GPU. As expected, a GPU does speed up backpropogation through the NN immensely as required for computing $z^*$ in \eqref{eq:gan_eq_rep}.

\begin{table}[!ht]
    \centering
    \includegraphics[width=5in]{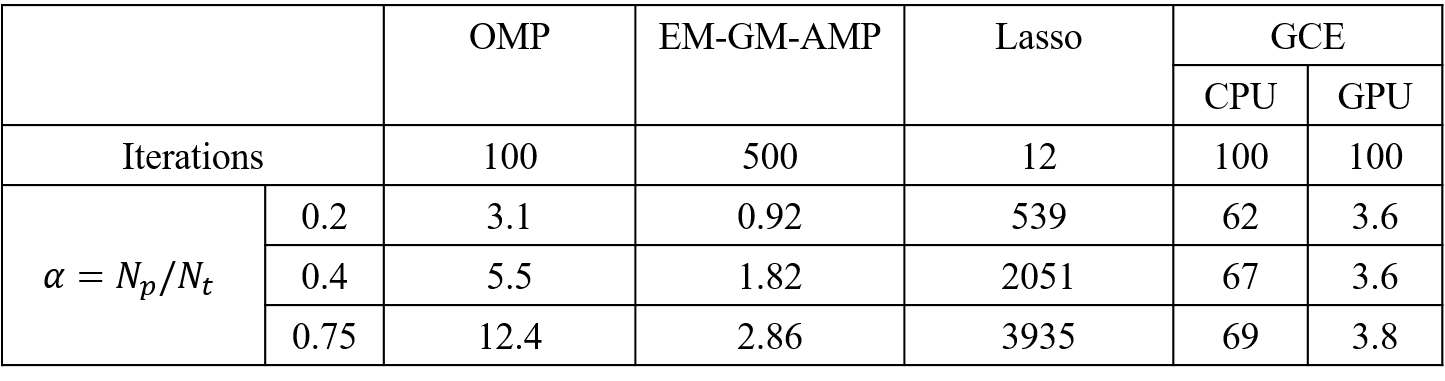}
    \caption{Comparison of execution time per iteration (in milliseconds) for OMP, Lasso, EM-GM-AMP and GCE on a single channel realization at an SNR of -10 dB.}
    \label{tab:execution_time}
\end{table}

The most important finding from Table \ref{tab:execution_time} is that the execution time of GCE is not noticeably affected by the increase in the number of pilot symbols, while the execution time of CS baselines increases with increase in $\alpha$. The gradient of \eqref{eq:gan_eq_rep} with respect to $z$ is given by
\begin{equation}
    \nabla_z f(\mathbf{y},\mathbf{AG}(z)) = 2 ( \mathbf{A}^T (\underline{\mathbf{y}} - \mathbf{A}\underline{\mathbf{G}}(z))\nabla_z \underline{\mathbf{G}}(z) + \lambda_{reg}z),  
\end{equation}
where each row of the matrix $\nabla_z \underline{\mathbf{G}}(z)$ is $d = 35$ dimensional. This involves only direct matrix multiplications of $\mathbf{A}$.
On the other hand, for OMP, one of the steps involves inverting columns of $\mathbf{A}_{\mathrm{sp}}$ having maximum inner product with $\underline{\mathbf{y}}$, whose complexity scales as $\mathcal{O}(N_{\mathrm{p}}^{\mathrm{m}})$ with $2\leq m < 3$. Similarly, the Lasso and EM-GM-AMP optimization problems have complexity scaling with $N_{\mathrm{p}}^{\mathrm{m}}$. Moreover, as explained in Section \ref{subsec:baselines}, Lasso involves solving an SOCP, hence takes much longer than the other algorithms. The impact of $N_\mathrm{p}$ on the execution time of GCE will only be seen at much higher values of $N_\mathrm{p}$, unlike the CS based algorithms for whom the impact of increasing $N_\mathrm{p}$ is immediately apparent\footnote{Each entry of a matrix multiplication can be computed in parallel, but is limited by the number of threads available on the CPU/GPU. Matrix inversion cannot be parallelized in the absence of its LU factorization.}. Note however that the complexity of computing $\nabla_z \underline{\mathbf{G}}(z)$ is quite high owing to the large number of weights $\theta_g$ in the trained generator $\mathbf{G}$, hence the execution time of GCE is comparable to OMP in the low pilot density regime.

\section{Conclusion}\label{sec:conc}
We presented a compressed sensing-based channel estimation approach using deep generative networks that achieves a significant performance gain over prior techniques for sparse signal recovery, when applied to CDL channel models. Notable aspects of this approach are that it does not require knowledge of the sparsifying basis of the channel and immensely reduces the number of pilots required to achieve the same NMSE as Lasso/OMP/EM-GM-AMP channel estimation, even in the case of one-bit quantized pilot measurements. Importantly, as a consequence of the gradient computation of \eqref{eq:gan_eq_rep} requiring only matrix multiplications, its execution time is approximately independent of the number of received pilot symbols $N_p$ when $N_\mathrm{p}$ is small. 

\section{Acknowledgements}
The authors would like to thank Nitin Myers for discussions on low resolution quantization and Shilpa Talwar, Nageen Himayat, Ariela Zeira at Intel for their invaluable support and technical advice and feedback.

\bibliographystyle{IEEEtran}
\bibliography{bibtex.bib}

\end{document}